\documentclass[prd,floatfix,nofootinbib]{revtex4}
\usepackage{amsmath}
\usepackage{amssymb}
\usepackage{ifpdf}
\usepackage{color}
\usepackage[pdftex]{graphicx}
\usepackage{fancyhdr}
\usepackage{extramarks}
\usepackage{wasysym}
\usepackage{pstricks}
\usepackage{bbm}
\usepackage{epsfig}
\usepackage{natbib}
\usepackage{chapterbib}
\usepackage{psfrag}
\usepackage[percent]{overpic}
\usepackage{multirow}
\usepackage{ifthen}
\usepackage{hyperref}
\usepackage{calc}
\usepackage{rotating}
\usepackage{latexsym}
\usepackage{subfigure,rotating,rotate}
\usepackage{color}
\usepackage{breakurl}

\input{superbsym}

\def\ze#1   {\ensuremath{\zeta_{#1}}\xspace}

\def\superb   {Super$B$\xspace}

\def\Bz   {\ensuremath{B_{z}}\xspace}

\def\CO2  {$\mathrm{CO}_2$\xspace}

\definecolor{light-gray}{gray}{0.7}

\newcommand{\CPV}{\ensuremath{C\!PV}\xspace}

\long\def\inst#1{\par\nobreak\kern 4pt\nobreak
    {\it #1}\par\vskip 10pt plus 3pt minus 3pt}

\newcommand{\beq}{\begin{equation}}
\newcommand{\eeq}{\end{equation}}
\def\be{\begin{equation}}
\def\bea{\begin{eqnarray}}
\def\eea{\end{eqnarray}}

\newcommand{\e}      [1]   { {\ensuremath{ \times 10^{ {#1} } }}}

\makeatletter
\DeclareRobustCommand\bfseries{%
  \not@math@alphabet\bfseries\mathbf
  \fontseries\bfdefault\selectfont\boldmath}

\makeatother

\makeatletter
\newcommand{\rulesandwich}[1]{%
  {\centerline{\rule[0.5ex]{0.9\linewidth}{2pt}}}\par\nopagebreak\vskip 0.5ex%
  \centering #1\par\nopagebreak\vskip 0.5ex%
  {\centerline{\rule[0.5ex]{0.9\linewidth}{2pt}}}\smallskip
}
\def\before@@par#1#2\@@par{#2#1\@@par}
\@latex@info{//// style change: modifying section title style}
\def\@hangfrom@section#1#2{#1#2}%
\def\section{%
  \@startsection%
  {section}%
  {1}%
  {\z@}%
  {0.8cm \@plus1ex \@minus .2ex}%
  {0.5cm}%
  {%
    \normalfont\small\larger\larger\bfseries\rulesandwich
  }%
}%
\def\@part[#1]#2{%
 \@ifnum{\c@secnumdepth >\m@ne}{%
        \refstepcounter{part}%
        \addcontentsline{toc}{part}{\thepart\hspace{1em}#1}%
 }{%
      \addcontentsline{toc}{part}{#1}%
 }%
 \begingroup
    \parindent \z@ \raggedright
    \interlinepenalty\@M
    \@ifnum{\c@secnumdepth >\m@ne}{%
      \Large \bf \partname~\thepart%
      \par\nobreak
    }{}%
    \normalfont\small\larger\larger\larger\larger\bfseries
    {\centerline{\rule[0.5ex]{0.9\linewidth}{2pt}}}\par\nopagebreak\vskip 1ex%
    \centering #2\par\nopagebreak\vskip 1ex%
    {\centerline{\rule[0.5ex]{0.9\linewidth}{2pt}}}%
    \markboth{}{}\par
 \endgroup
   \nobreak
   \vskip 3ex
   \@afterheading
}%
\def\@spart#1{{\parindent \z@ \raggedright
    \interlinepenalty\@M
    \normalfont\small\larger\larger\larger\larger\bfseries
    {\centerline{\rule[0.5ex]{0.9\linewidth}{2pt}}}\par\nopagebreak\vskip 0.5ex%
    \centering #1\par\nopagebreak\vskip 0.5ex%
    {\centerline{\rule[0.5ex]{0.9\linewidth}{2pt}}}\par
  }
  \nobreak
  \vskip 3ex
  \@afterheading}
\makeatother

\def\LALRNUM{LAL-11-200}
\def\INFNRNUM{INFN/AE\_11/1}
\def\SLACRNUM{SLAC-R-14548}

\begin{document}

\pagestyle{empty}

\onecolumngrid
\begin{flushright} 
\INFNRNUM, \LALRNUM, \SLACRNUM, MZ-TH/11-25 
\end{flushright} 
\begin{center}
{\boldmath{
\phantom{\Huge{I}}
\phantom{\Huge{I}}
%\phantom{\Huge{I}}
%\phantom{\Huge{I}}
%\phantom{\Huge{I}}
%\phantom{\Huge{I}}
%\phantom{\Huge{I}}
%\phantom{\Huge{I}}
%\phantom{\Huge{I}}
%\phantom{\Huge{I}}

{\huge\bf
{\vskip 12pt
The impact of \superb on flavour physics}}\\
{\vskip 12pt \Large\bf July 1, 2011}\\
}
}
\end{center}

\bigskip
% Abstract
\begin{center}
{\large \bf Abstract}
\end{center}
This report provides a succinct summary of the physics programme of \superb, and describes
that potential in the context of experiments making measurements in flavour physics over the
next 10 to 20 years. Detailed comparisons are made with \belletwo and \lhcb, the 
other $B$ physics experiments that will run in this decade. SuperB will play a 
crucial role in defining the landscape of flavour physics over the next 20 years. 
$\,$

%
% Author list for the comparison document (in case we need it)
%
\author{B.~Meadows}
\affiliation{University of Cincinnati, Cincinnati, Ohio 45221, USA}

\author{M.~Blanke}
\affiliation{Laboratory for Elementary Particle Physics, Cornell University, Ithaca, NY 14850, USA}

\author{A.~Stocchi}
\affiliation{Laboratoire de l'Acc\'{e}l\'{e}rateur Lin\'{e}aire, IN2P3/CNRS et Universit\'e de Paris-Sud XI,
Centre Scientifique d'Orsay, F-91898 Orsay Cedex, France}

\author{A.~Drutskoy}
\affiliation{Institute for Theoretical and Experimental Physics, B. Cheremushkinskaya 25, 117218 Moscow, Russia}

\author{A.~Cervelli}
\author{M.~Giorgi}
\author{A.~Lusiani}
\author{A.~Perez}
\author{J.~Walsh}
\affiliation{Universit\`a di Pisa, Dipartimento di Fisica, Scuola Normale Superiore
and INFN, Pisa, Italy}

\author{T.~Hurth}\altaffiliation{Also affiliated with CERN, CH-1211 Geneve 23, Switzerland}
\affiliation{Institute for Physics, Johannes Gutenberg-University, 55099 Mainz, Germany}

\author{A.~Bevan}
\affiliation{Queen Mary, University of London, E1 4NS, United Kingdom}

\author{L.~Silvestrini}
\affiliation{INFN Roma, I-00185 Roma, Italy}

\author{M.~Ciuchini}
\affiliation{INFN Sezione di Roma Tre, I-00146 Roma, Italy}

\author{C.~Tarantino}
\affiliation{Dipartimento di Fisica, Universit\`a Roma Tre and INFN Sezione di Roma Tre I-00146 Roma, Italy}

\maketitle

$\,$
\vfil
\clearpage
\newpage

%\vskip -1.8cm
\tableofcontents
\onecolumngrid
%\twocolumngrid
\hbox{}

%\vfill
\clearpage

\pagestyle{fancyplain}

\fancyfoot{} % clear all fields
\fancyfoot[LE,RO]{\it{The impact of \superb on flavour physics}}
\fancyhead{} % clear all fields

\renewcommand{\sectionmark}[1]%
                  {\markright{#1}}
\rhead[\fancyplain{}{\bf Introduction}]%
      {\fancyplain{}{\bf\thepage}}

\graphicspath{{Introduction/}{Introduction/}}
\section{Introduction}
\label{sec:introduction}

% introduction to the superb physics programme, setting the stage and providing recent
% references to the white papers, Valencia and CDR.

\superb is an approved high luminosity \epem collider intended to search 
for indirect and some direct signs of new physics (NP) at low energy, while at the 
same time, enabling precision tests of the Standard Model (SM).  This experiment
will be built at a new laboratory on the Tor Vergata campus near Rome, Italy
named after Nicola Cabibbo. The
project has been described in a Conceptual Design Report~\cite{Bona:2007qt}, 
and more recently by a set of three white papers on the accelerator~\cite{Biagini:2010cc}, 
detector~\cite{Grauges:2010fi}, and physics programme~\cite{O'Leary:2010af}.

The main focus of the physics programme 
rests in the study of so-called {\em Golden Modes}, these
are decay channels that provide access to measurements of theoretically clean 
observables that can provide both stringent constraints on models of 
NP, and precision tests of the SM.  A number of ancillary measurements
that remain important include those with observables that may not be 
theoretically clean, and those that can be used to provide stringent 
constraints on the SM but are not sensitive to NP. 
The remainder of this section introduces \superb before discussing
the golden modes for \superb, precision CKM measurement modes, 
and an outline of the rest of this report.

\subsection{\superb in a nutshell}

\superb will accumulate 75\invab of data over a period of five years of 
nominal data taking at the \FourS.  In addition to operating at the 
centre of mass energy of the \FourS, this experiment will also
run at other energies ranging from charm threshold, at the $\psi(3770)$, to 
the \FiveS.  \superb will be constructed and commissioned with an 
engineering run by the end of \builtby, and nominal 
data taking expected to commence in \startnominalrunning. Thus
the physics potential of \superb discussed here is expected to
re-define the flavour physics landscape by \fiveyears.

\subsection{Golden modes for \superb}
\label{sec:introduction:goldenmodes}

In the indirect search for NP using flavour-changing processes, there are two necessary ingredients (i) a
decay channel that can be experimentally studied to a sufficient precision to identify a deviation
from the SM, and (ii) sufficient theoretical understanding of the SM calculation to validate that any
inconsistency between theory and data is a real NP effect, as opposed to uncontrolled or underestimated
theoretical uncertainties. However, high sensitivity to NP also means that a decay channel allows for
significant deviations from the SM within NP models.  We will analyse this aspect by using a variety
of presently popular NP benchmark models in Section~\ref{sec:experimentalreach:interplay}. 
Direct searches, such as those for a light Higgs boson or Dark Matter candidates,
are more straightforward, where discovery of a new particle that
doesn't match any SM expectation would be a clean indication of NP.

Based on this premise, we list the Golden channels/observables relevant for \superb
in the following.  

\begin{description}
 \item{\bf $\tau$ Physics}
\begin{itemize}
  \item Lepton flavour violation in tau decays: $\tau \to \mu \gamma$ and $\tau \to 3\ell$
            as specific examples of the experimental reach.  
\end{itemize}
\end{description}

\begin{description}
 \item{\bf $B_{u,d}$ Physics}
\begin{itemize}
  \item $B^{+}\to \tau^{+} \nu$ and $B^{+}\to \mu^{+} \nu$
  \item $B^{+}\to K^{(*)+} \nu \overline{\nu}$
  \item $b\to s \gamma$
  \item $b\to s \ell \ell$
  \item $\Delta S$ measurements, in particular $B^0\to\eta^\prime \KS$, and $S$ in $B\to \KS\pi^0\gamma$.
\end{itemize}
\end{description}

\begin{description}
 \item{\bf $B_s$ Physics}
\begin{itemize}
  \item Semi-leptonic \CP asymmetry $A_{SL}^s$. 
  \item $B_s\to \gamma \gamma$.
\end{itemize}
\end{description}

\begin{description}
 \item{\bf Charm Physics}
\begin{itemize}
  \item mixing parameters and \CP violation. % $x$ and $y$
%  \item $y_{\CP}$
\end{itemize}
\end{description}

\begin{description}
 \item{\bf Other Physics}
\begin{itemize}
  \item Precision measurement of $\sin^2 \theta_{W}$ at $\sqrt{s} = 10.58$ GeV/$c^2$.
  \item Direct searches for non-standard light Higgs bosons, Dark Matter and Dark Forces
\end{itemize}
\end{description}

Each of these groups of golden measurements is described briefly 
in the following text.  

\begin{description}
\item{\bf $\tau$ Physics}
Since Lepton Flavor Violation (LFV) is severely suppressed in the SM,
LFV $\tau$ decays are especially clean and unambiguous
experimental probes for NP effects. \superb can experimentally access
$\tau$ LFV decay rates over 100 times smaller than \babar for the most
clean channels (e.g.\ $\tau\to 3\ell$), and over 10 times smaller for
other modes such as $\tau \to \ell \gamma$ that have irreducible
backgrounds. In all cases, the \superb polarised electron beam
provides additional advantages to determine the properties of the LFV
interaction from the polarization-dependent angular distribution of
the $\tau$ decay products, and to improve the selection for specific
NP models.
This gives \superb a distinct advantage over other experiments.

The electron beam polarization provides \superb with even more
significant advantages for measuring \CP violation in $\tau$ decay and
for measuring the fundamental $\tau$ properties corresponding to the
$g{-}2$ and EDM form factors from the angular distribution of the
$\tau$ decay products. In all three cases \superb will extend
considerably the present experimental physics reach, providing
sensitivity to some specific NP models and a first
statistically significant measurement of the $\tau$ $g{-}2$ form
factor.

This document concentrates only on the $\tau \to \ell \gamma$ and
$\tau\to 3\ell$ channels as specific examples to illustrate the
physics reach of \superb in this area, but one should not forget that
\superb will provide the same advantages to extend the 
experimental reach for all other $\tau$ LFV modes, some of which can
be most sensitive to specific NP models.

\item{\bf $B_{u,d}$ Physics} The main goals of the \B Physics
  programme at \superb cover observables measured using the decays of
  $B_u$ and $B_d$ mesons, and, with a smaller collected data sample,
  of $B_s$ mesons.  NP is expected to affect the predictions computed
  in the SM framework for many observables.  Some of the theoretically
  cleanest observables are related to final states with $\nu$'s or
  many neutral particles, and these can only be studied in an \epem
  environment, where \superb will have the largest data sample
  available.  Modes of particular interest include $B\to\ell \nu$,
  which can be used to constrain charged Higgs particles, and $B\to
  K^{(*)} \nu\bar\nu$, which is highly sensitive to Z penguin and
  other electroweak penguin effects.  The $b\to s\gamma$ and $b\to
  s\ell\ell$ transitions can be measured inclusively and exclusively
  at \superb.  In contrast it is only possible to exclusively
  reconstruct these modes in a hadronic environment. Golden
  observables at \superb include the various kinematical and angular
  distributions, as well as \CP and isospin asymmetries. For the
  branching ratios, it is important to perform both inclusive and
  exclusive measurements to ensure that results obtained from
  theoretical interpretation of the data are in agreement, and
  similarly it is important to measure both electron and muon final
  states in $b\to s\ell\ell$ in order to constrain all of the
  available NP sensitive observables.

The focus on time-dependent \CP violation measurements is twofold,
(i) constraining NP in loop (penguin) and tree processes and (ii) performing precision
CKM constraints.  Most of the $B_d$ decays of interest can
only be measured in an $e^+e^-$ environment: \superb will measure CP asymmetries
in a wide variety of hadronic $b\to s$ penguin modes, especially in the most precisely determined
penguin channel $\B^0\to \eta^\prime K^0$. Moreover, \superb will measure the photon
polarisation via the mixing-induced asymmetry in the $B^0\to \KS\piz\gamma$ mode. 
Other important channels of this kind are decays such as $B\to J/\psi \piz$, which is needed
to constrain the theoretical
uncertainties on the tree level $b\to c\overline{c}s$ $\beta$ measurement.  Furthermore, final states
with $K^0_L$ mesons provide useful cross checks of the \KS decays (with 
opposite \CP eigenvalues).  It would be challenging to try and reconstruct 
these important $B_{u,d}$ modes in a hadronic environment: at \lhcb the 
reconstruction efficiencies are reduced for final states containing several
neutral particles and for analyses where the $B$ decay vertex must be 
determined from a $K_s^0$ meson. 
However, \lhcb can perform complementary measurements using $B_s$ decay 
modes such as 
$B_s \to J/\psi \phi$, $B_s \to \phi \gamma$ and $B_s \to \phi\phi$, as discussed below in Section~\ref{sec:otherexperiments:lhcb}.

\item{\bf $B_s$ Physics}
An expected data sample of approximately 1\invab collected at the
$\Upsilon(5S)$ resonance will provide a large sample of $B_s$ decays.
This sample will allow comprehensive studies of the decay rates of the $B_s$
that are comparable in precision to the currently available results
for $B_{u,d}$ mesons.  Additionally, two important contributions to
$B_s$ physics that can be made by \superb are a precision measurement
of the semi-leptonic asymmetry $A_{SL}^s$, which is complementary to
the measurement of $\beta_s$ for searching for NP in $B_s$ mixing, and
a measurement of $B_s\to \gamma\gamma$. 

Hadron experiments are able to measure $A_{SL}^s$, however the ultimate precision
attainable will be limited by knowledge of systematic uncertainties related to 
charge asymmetries in the data.  In order to overcome this limitation, it is 
currently preferred that one measures an asymmetry difference 
$\Delta A^{s,d} = A_{SL}^s - A_{SL}^d$~\cite{lhcbasym}.
The measurement of $A_{SL}^s$ in an \epem environment will be useful as 
this will be subject to a 
different set of systematic uncertainties that will ultimately limit the 
precision, however it is expected that one can measure $A_{SL}^s$ directly without
having to resort to an asymmetry difference in order to overcome the lack
of knowledge of charge asymmetry in data.

In general the decay $B_s\to \gamma\gamma$ can be 
affected in a similar way to $B_{u,d}\to X\gamma$ by the presence of NP
(for example see ~\cite{PhysRevD.70.035008}), making this an important channel to measure.
In some models however the $B_s$ and $B_{u,d}$ decays are uncorrelated and
in those cases the $B_s$ channel may have better sensitivity~\cite{Huo:2003cj,Aranda:2010qc}
to NP.

 \item{\bf Charm Physics} The main goals of charm physics currently
   are to perform precision measurements of mixing in the charm
   system, and to search for possible \CP violation in the up-quark
   sector.  A distinct advantage that one has at \superb in this area
   is that in addition to large samples of relatively clean events
   accumulated at the \FourS, one can utilise data collected at the
   $\psi(3770)$, in particular to constrain strong phases and Dalitz
   models required for charm mixing measurements and the determination
   of the Unitarity triangle angle $\gamma$ using the standard GGSZ method, to
   reduce model uncertainties.  Data collected at the $\psi(3770)$
   provide an almost pure ``$D$ meson beam'' that can be used to
   perform precision measurements that can be used to reduce 
   systematic uncertainties for measurements made at
   the \FourS or in a hadronic environment.  In
   some cases, results from such a run can compete with results from
   the \FourS data reported here.  Therefore, depending on the run
   schedule, some results could become available at a similar level of
   precision earlier in the \superb cycle.  Time-dependent \CP
   asymmetry studies in $D$ decays have recently been proposed in 
   Ref.~\cite{Bevan:2011up}, and studies are in progress to 
   understand the potential of such measurements at \superb.

 \item{\bf Other Physics}
A precise measurement of the left-right asymmetries of $e^+e^-\to \ell^+\ell^-,\,b\bar b$
can be performed at \superb thanks to the polarized electron beam.
This allows for a determination of $\sin^2\theta_W$ at the \FourS with precision 
comparable to the LEP/SLC measurement at the $Z$ pole, which is of interest for
two particular reasons (i) a measurement at the \FourS is theoretically
cleaner than one at the $Z$ (no $b$ fragmentation uncertainties), and 
(ii) there is no measurement at this particular energy which is in a
region where $\sin^2\theta_W$ is changing rapidly.  This result 
can be fed into precision electroweak fits along with existing and other planned
measurements.  The JLab experiment QWeak~\cite{qweak} 
will improve the precision of the $\sin^2\theta_W$ at a lower $\sqrt{s}$ 
than \superb, and there is a proposed
experiment called MESA, at Mainz, to perform a measurement at an even smaller energy
scale.

The previously mentioned golden channels and observables are focused on
observation of NP effects through indirect means.  In addition to SM
spectroscopy, \superb can also search for light dark matter, light 
Higgs particles that could exist in multi-Higgs scenarios beyond the SM,
and to also search for evidence of Dark Forces. 
\end{description}

\subsection{Precision CKM measurement modes}
\label{sec:introduction:calibrationmodes}

In addition to the aforementioned Golden modes, there is a list of precision measurements
that can be used to determine the CKM parameters at the percent level and
to test the SM by checking the compatibility of experimental results.
Current data indicate a number of inconsistencies at the level 
of 2--3$\sigma$ in these constraints which deserve to be further investigated.
Yet, even if no NP signal will emerge from these measurements, a precise determination
of the CKM parameters is a crucial ingredient to improve the SM prediction, and hence
the NP sensitivity, of several flavour observables.

The precision CKM measurements include
\begin{itemize}
  \item \Vub (inclusive and exclusive measurements)
  \item \Vcb (inclusive and exclusive measurements)
  \item $\alpha$ from $b\to u\overline{u}d$ transitions
  \item $\beta$ from $b\to c\overline{c}s$ transitions
  \item $\gamma$ from $B \to DK$ transitions
\end{itemize}

Data collected at or just above the $\psi(3770)$ resonance can
be used to measure \Vcd and \Vcs, and one can perform a precision measurement of
\Vus using $\tau$ decays.
\superb is the only experiment that can complete this set of measurements in order
to perform direct and indirect precision tests of the CKM mechanism.

\subsection{Other measurements}
\label{sec:introduction:other}

There is a vast potential to perform other measurements that are not classified as either a {\em golden mode}
or a {\em precision CKM measurement mode}.  These include hundreds of possible
measurements of $B$, $D$, $\tau$, $\Upsilon$, and $\psi(3770)$ decays as well as studies of ISR processes
and both conventional and exotic spectroscopy not listed above.  A partial description of these 
possibilities is listed in Refs.~\cite{O'Leary:2010af,Bona:2007qt,Hitlin:2008gf},
and will be discussed in the context of the result of existing experiments in the forthcoming Physics of 
the B Factories book currently in preparation~\cite{PBF}.  \superb will integrate 150 (75) times the 
data of \babar and \belle enabling a significant improvement in precision of these other measurements.
This area of the programme will not be discussed further here.

\subsection{The remainder of this document}
\label{sec:introduction:thisdocument}

The remainder of this document includes a brief summary of flavour physics experiments related to 
the \superb programme (Section~\ref{sec:otherexperiments}).  These include the flavour physics experiments
\belletwo and \lhcb that have some measurements in common with \superb, plus those flavour experiments that provide complementary information, or for 
which \superb measurements themselves will be useful.

Section~\ref{sec:experimentalreach} summarises the experimental reach of \superb based on the 
physics studies that have been previously described in 
Refs.~\cite{O'Leary:2010af,Bona:2007qt,Hitlin:2008gf}.  Where appropriate,
these measurements are compared with the expectations of other experiments.  
Section~\ref{sec:experimentalreach} also revisits the issue of interplay between measurements
to highlight the complementarity of different searches for NP in the context
of the time-scale of the first five years of data taking. 
Finally Section~\ref{sec:summary} provides a succinct summary
of the physics programme for \superb in the context of the experimental landscape circa \fiveyears.

%
% Other experiments
%
\renewcommand{\sectionmark}[1]%
                  {\markright{#1}}
\rhead[\fancyplain{}{\bf Other experiments}]%
      {\fancyplain{}{\bf\thepage}}

\graphicspath{{Experiments/}{Experiments/}}
\section{Other experiments}
\label{sec:otherexperiments}

One purpose of this report is to put \superb in the
context of current and next-generation flavour experiments.
Indeed, one should not forget that there are other important
flavour experiments that can be of interest when interpreting
results from \superb.  Some of the measurements of these
other experiments provide complementary constraints on
new physics that could have an impact on the \superb physics
programme, and in return some of the measurements 
possible at \superb will feed back into the physics programme
of those other experiments.

\begin{figure*}[!ht]
\begin{center}
\resizebox{12.cm}{!}{
\includegraphics{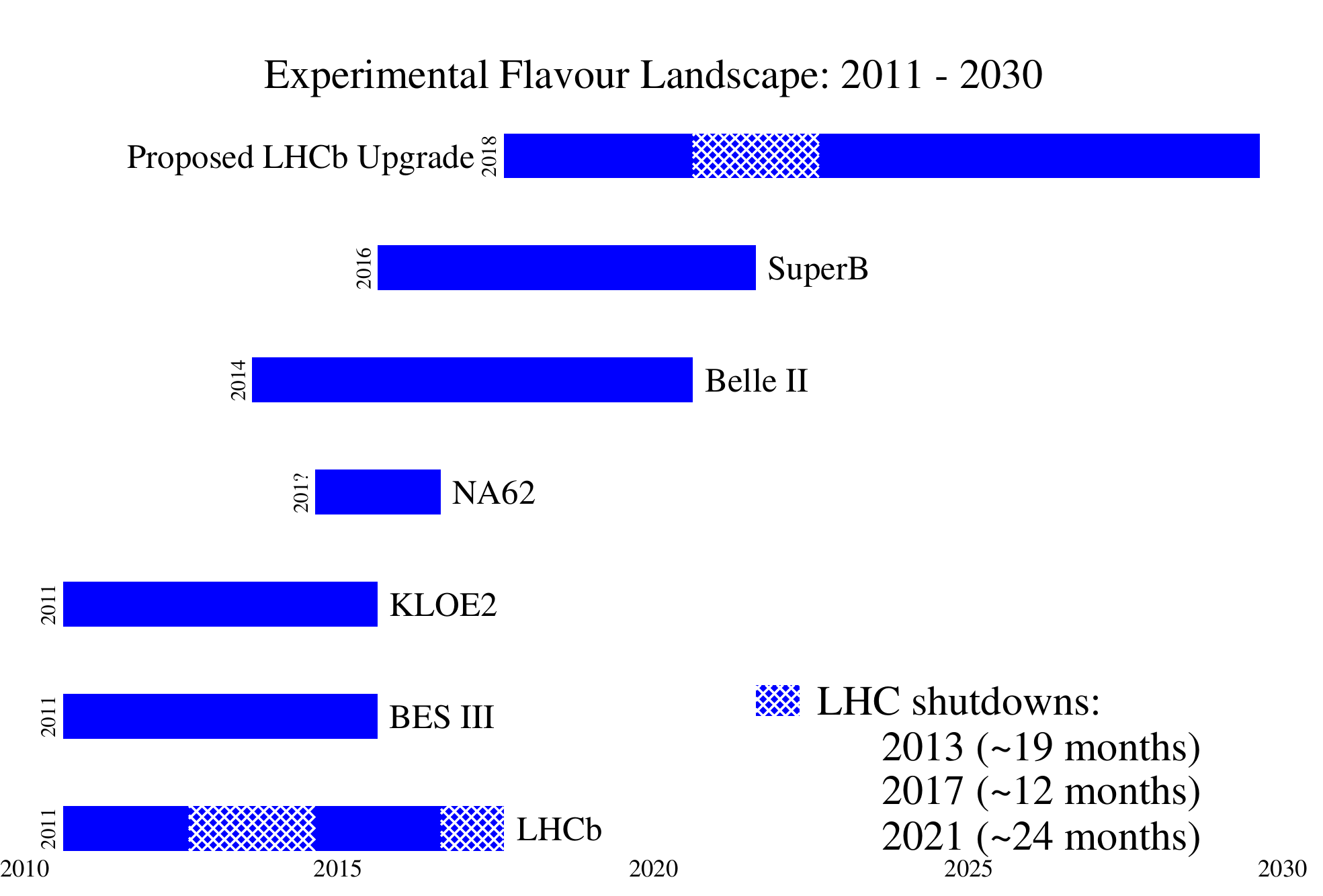}
}
\caption{Time scales of data taking anticipated for existing, approved, and proposed 
flavour physics experiments discussed in this document.  One can see that \lhcb, \besthree
and NA62 will have completed their currently defined run programmes prior to the \superb and \belletwo.
Any upgrade of \lhcb would achieve final results sometime after the Super Flavour Factories
will have achieved their goals.
}\label{fig:otherexpts:timescales}
\end{center}
\end{figure*}

There are two other experiments that will focus on $B$ physics in this
decade: \belletwo, currently under construction at KEK in Tsukuba,
Japan, and \lhcb, currently operating at CERN in Geneva.
Additionally, there are several flavour physics experiments that are
relevant for the physics programme of \superb in terms of providing
complementary constraints on new physics, and in terms of requiring
input from \superb measurements.  These other experiments are
\besthree, \comet, KLOE 2, \koto, \meg, \mutoe, and NA62.  A planned upgrade
of \lhcb may start taking data shortly after \superb, but take longer
to complete its physics programme.  An estimate of the relative time
scales for data taking for these experiments are shown in
Figure~\ref{fig:otherexpts:timescales}.

The remainder of this section describes the relationship between
\superb and these other experiments in the context of the global
picture of constraining new physics.

\subsection{\belletwo}

The Belle II experiment, an $e^+e^-$ experiment running at the \FourS,
has many similarities with SuperB, and there is a large overlap in
their respective physics programmes.  There are several important
differences between these two Super Flavour Factories:

\begin{itemize}
  \item On the time-scale that \superb will accumulate 75\invab, it is
    expected that \belletwo will be able to integrate 50\invab at the
    \FourS.  One would then expect that \superb will outperform
    \belletwo in terms of precision by about 20\% in general for
    measurements that are not limited by systematic uncertainties.
    Where measurements will be limited by systematic uncertainties,
    the expectation of the ultimate precision reached depends on
    assumptions that have entered into extrapolations, and differences
    between experiments should be interpreted as a possible range of
    the ultimate precision attainable.

  \item \superb can run at energies below the \TwoS, and will run at
    both the \OneS and $\psi(3770)$.  This provides \superb with 
    a significantly broader physics programme through the direct
    searches for light scalar mesons (Higgs and Dark Matter), tests
    of lepton universality, searches for Dark Forces, and so on.
    Measurements at charm threshold will feed back into the \B physics
    programme of all flavour experiments: for example reducing 
    model uncertainties in the measurement of the Unitarity triangle
    angle $\gamma$, and improving the precision of charm mixing
    measurements.  Other charm threshold measurements will also 
    enable lattice QCD to be tested more precisely in a regime
    where calculations are better understood.  This will impact
    upon the corresponding work in $B$ decays.

  \item The electron beam at \superb will be polarised to at least
    80\% for running at the \FourS.  The polarized electron beam
    provides \superb with the ability to perform precision electroweak 
    studies in an energy regime free from hadronic uncertainties related to $b$
    fragmentation that otherwise limit the interpretation of SLC/LEP 
    measurements, and also provides an additional kinematic
    variable to support background fighting techniques in 
    rare and Lepton Flavour Violating (LFV) $\tau$ decay studies.
    The benefits of polarisation for $\tau$ LFV are model dependent.
\end{itemize}
More details on the \belletwo experiment and physics programme can be found
in Ref.~\cite{Aushev:2010bq}.

\subsection{\besthree}
The physics programme of BES III, an $e^+e^-$ experiment running at
charm threshold, will overlap significantly with the SuperB
measurements made at charm threshold.  The anticipated data sample to
be accumulated at the $\psi(3770)$ by \besthree is 10\invfb by the
middle of the this decade~\cite{haiboLiElba}.  In comparison, \superb
expects to be able to accumulate 500\invfb with several months of
running\footnote{It is likely the \superb threshold run will take
  place after the machine has been optimised for \FourS operation.
  This would be a few years into the data taking life of \superb.}.
Thus \superb will be able to make significant improvements on the
precision of measurements made by \besthree: in general SuperB results
will have statistical uncertainties reduced by a factor of seven for
all observables where these experimental programmes overlap.  As
\superb will take data after \besthree we expect that there will be a
natural evolution in the understanding of flavour physics at charm threshold
from one experiment to the other.  More details on the \besthree
experiment and physics programme can be found in Ref.~\cite{Asner:2008nq}.

\subsection{\lhcb (and planned upgrade)}
\label{sec:otherexperiments:lhcb}

Although \lhcb is a $B$ physics experiment, the starkly different
running environment of a hadronic machine dictates a physics programme
that has only moderate overlap with the \superb programme.  LHCb
benefits from a high $b$ cross section and all types of $B$-hadrons
are produced in the $pp$ collisions, including a large number of $B_s$
mesons.  The most striking outcome of any comparison between \superb and \lhcb is that the
strengths of the two experiments are largely complementary. For example, the large
boost of the $B$ hadrons produced at LHC allows for studies of the oscillations
and mixing-induced CP violation of $B_s$ mesons, while most of the measurements
that constitute the primary physics motivation for \superb cannot be performed in the
hadronic environment with its higher backgrounds and trigger difficulties (as previously
mentioned in Section~\ref{sec:introduction:goldenmodes}). 

Clearly, there is some uncertainty as to when one can expect results with full integrated
luminosity from \lhcb and \lhcb upgrade. Presently, the following 
shutdowns at CERN are planned over the next decade~\cite{bairdforaz} 
\begin{itemize}
  \item $2013-2014$: A 15-19 month shutdown (duration not yet decided) for upgrades
        to the LHC, where test-beams, the PS and SPS are expected to be unavailable 
        during the shutdown as manpower will be redirected to LHC upgrade work.
  \item $2017$: A shutdown of some length (to be confirmed) for the ATLAS and 
   CMS experiments to install pixel detector upgrades.
  \item $2021$: Likely date to commence a shutdown for a future luminosity 
    upgrade to the LHC.  The length of this shutdown is unknown,
    by it is probably between 18 and 24 months.
\end{itemize}

Based on the Chamonix Conference presentations in January
2011 and \lhcb upgrade LOI~\cite{SergioBertolucci,lhcbupgradeloi}, we expect 
that \lhcb will accumulate 5\invfb by the end of its data taking life\footnote{\lhcb is limited to
  a maximum integrated luminosity recorded of 1\invfb per year,
  integrating 5\invfb by 2017, instead of the 2\invfb/yr originally
  expected.}, and that the \lhcb upgrade accumulate 50\invfb.
%
%\footnote{The \lhcb upgrade
%  is limited to a maximum integrated luminosity recorded of 5\invfb
%  per year.  If one assumes an instantaneous transition from \lhcb to
%  the upgrade and a 24 month shutdown for installation of general
%  purpose detector upgrades circa 2021, the end of the life of the
%  \lhcb upgrade will be 2030.  It should be noted that this estimate
%  is rather uncertain at this time.  The reasons for this include the
%  fact that the upgrade is not yet approved at CERN or among the
%  member states, and the time scale is dependent upon on a number of
%  external factors such as the \lhc machine performance and upgrade
%  schedule, and the general purpose detector (ATLAS/CMS) upgrade
%  schedules.}  
%to a total data sample of 50\invfb.
%
%Predicting accurately the time line of an upgraded \lhcb experiment is
%very difficult, as only a Letter of Intent exists so far, but this
%proposal specifies a start date of 2018.  
The experiment will be able
to collect a maximum of 5\invfb of data per year starting in 
2018~\cite{lhcbupgradeloi}, so 10 years of
running will be required to reach the goal of 50\invfb.  There will
also be expected shutdowns of LHC for upgrades of the accelerator and
other experiments (as indicated above), 
but the timing and duration of those shutdowns has
not yet been determined.  In any case, a {\em possible} time line for an
upgraded \lhcb, and several other flavour experiments, is given in
Figure~\ref{fig:otherexpts:timescales}.  Despite the uncertainty in
mapping out the time scales, we think it is fairly safe to state that
by and large the results from \lhcb will precede those of \superb,
which in turn will produce its results well in advance of the
conclusion of an upgraded \lhcb experiment.  While the majority of the
main motivational measurements made by these experiments are distinct,
there will be synergy in cross-calibration of the few modes that are
in common. In addition, any potential upgrade of \lhcb will benefit
from the measurements made at \superb. More details on the \lhcb
experiment and physics programme can be found in
Ref.~\cite{Adeva:2009ny}.

The following assumes an integrated luminosity of 5\invfb, achieved by 2017.
\begin{description}
   \item {\bf $B_s^0$ physics}: The \lhcb programme in this area is largely complementary to that of
    \superb.  Any NP discovery in the $B_s$ sector at \lhcb would naturally motivate complementary
    searches for NP at \superb, and vice versa. Golden $B_s$ observables at \lhcb include
    \begin{description}
      \item {$B_s^0 \to \mu^+\mu^-$}: CDF has recently produced a
        preliminary result of $BR(B_s\to \mu^+\mu^-) =
        (1.8^{+1.1}_{-0.9})\times 10^{-9}$ (with a corresponding upper
        limit of $< 4 \times 10^{-8}$ at 95\% C.L.)~\cite{cdf:bsmumu},
        while LHCb and CMS have now produced a combined LHC result where the branching
        ratio is limited to be $<1.5 \times 10^{-8}$ (95\%
        C.L.)~\cite{LHCbBsMuMu}. The SM expectation for this
        channel is $3.6\times 10^{-9}$, and has about a 10\%
        theoretical uncertainty.  It is expected that \lhcb with
        5\invfb of data will reach a precision of 30\% on the
        branching ratio of this mode assuming a SM
        rate~\cite{lhcbupgradeloi}.  NP can appear in loops for this
        decay, which is of relevance when one considers the interplay
        between this observable and those from \superb.

      \item {$2\beta_{s}$
        from $\B_s^0 \to J/\psi \phi$}:  The current precision on this observable from
       the Tevatron is $\sim 0.35$, and LHCb has recently presented a result with a precision
       of $\pm 0.19$.  It is expected that \lhcb will be able to measure this quantity
       to a precision of 0.019 with 5\invfb, where theoretical uncertainties are of the order of
       0.003~\cite{lhcbupgradeloi}.
      \item{$B_s\to \phi\gamma$}: \lhcb can measure the time-dependent \CP asymmetry in this channel which
       is sensitive to the presence of new right-handed currents. A precision of $~0.07$ on
       $S$ is expected~\cite{lhcbupgradeloi}.
   \end{description}

  \item {\bf The measurement of $\gamma$}:  This is a precision CKM measurement that is useful
    for over constraining the CKM mechanism. \lhcb should have measured $\gamma$ to a precision 
    of $4^\circ$ by the time \superb starts taking data.  \superb will reach an ultimate 
    precision of $\sim 1^\circ$ on $\gamma$ for any given channel.

   \item {\bf $\sin(2\beta)$ from $J/\psi\KS$}: This is a precision CKM measurement that is useful
     for over constraining the CKM mechanism. The \B Factories have measured $\beta$ to a
     precision of $0.8^\circ$, and it is expected that \lhcb will reduce this uncertainty to
     $0.5^\circ$. \superb will be able to improve on this and measure $\beta$ with a precision
     of $0.1^\circ$.

   \item {\bf $\B \to K^{(*)} \mu\mu$}: 
     \lhcb is expected to accumulate approximately 8,000 events in the
     channel $B^0 \to K^{*0}\mu^+\mu^-$~\cite{LHCbKstMuMu}.  This
     large data sample will allow precise measurements of the lepton
     forward-backward asymmetry, 
     and other angular observables sensitive to
     the presence of NP (for example see~\cite{Egede:2010zc,Egede:2008uy}).
     \lhcb will have significantly lower
     efficiency for the $B^0\to K^{*0}e^+e^-$ channel and will
     accumulate perhaps 400 events~\cite{LHCbkstee,WilkinsonElba}.
     \superb will select a larger sample of $B \to K^{*}\mu^+\mu^-$
     decays ($\sim 10,000 - 15,000$) and will accumulate a roughly
     equal number of $B\to K^*e^+e^-$ decays, plus a large sample of
     fully inclusive $B\to X_s\ell^+\ell^-$.  We further comment on
     $b\to s\ell^+\ell^-$ physics in
     Section~\ref{sec:experimentalreach}.
     
   \item {\bf Lepton Flavour Violation}: Preliminary studies show that LFV searches by \lhcb will at
     best match existing limits from \babar and \belle~\cite{GuzabProcs,Abe:2007ev,Lees:2010ez}.

   \item {\bf Charm mixing}: 
   %% {\red To understand better the relationship between charm threshold and
   %%  4S running.}
   In Ref.~\cite{O'Leary:2010af} it was concluded that, with 75~\invab 
at the $\Upsilon(4S)$, \superb could allow
measurements of $x$ and $y$ with a precision of a few parts
in $10^{-4}$.  \dzdzb asymmetries in these parameters would allow
estimation of the complex \CPV parameter
$q/p$ with uncertainties $\sim 1$\% and $\sim 1.5^{\circ}$, in 
magnitude and phase, respectively\footnote{%
The observables are the magnitude and phase of
$\lambda_f=\frac{q}{p}\frac{\bar A_f}{A_f}$ where $A_f(\bar A_f)$ are
the amplitudes for decays of $\Dz(\Dzb)\to f$.  All data have, so far,
been analyzed in terms of models (or measurements) for $A_f(\bar A_f)$
with specific hypotheses for the phase of $\frac{\bar A_f}{A_f}$.  In
this sense, the magnitude and phase of $\frac{q}{p}$ have been
regarded by the HFAG as observables, though they do, of course, depend
on the models used for each decay mode $f$.}.
Measurements of $|q/p|$ in a broad mix of individual channels with 
a precision of a few percent are expected that could reveal 
whether \CPV comes from direct or indirect sources.

The most precise results are expected to come from time-dependent 
Dalitz plot (TDDP) analyses of decays of \Dz to self-conjugate final 
states ($\KS\pip\pim$, $\KS\Kp\Km$, etc.)
\footnote{%
The precision from the TDDP analyses is systematically limited by
knowledge of the proper decay amplitude model to use to describe the
Dalitz plot populations.  Strong phase measurements made at charm 
threshold should substantially reduce this limitation for \lhcb, 
\superb and \belletwo.   A 500~\invfb charm threshold run could
reduce this limitation, and a time-dependent analysis of mixing could
add further, independent information on mixing and \CPV.  Estimates in
Table~\ref{tbl:experimentalreach:golden} assume this uncertainty can be 
ignored.}  %
and, for $y_{CP}$, from lifetime measurements for decays to \CP eigenstates
($\Kp\Km$, $\pip\pim$, etc.).

Statistical uncertainties in $x$, $y$, $y_{CP}$, $|q/p|$ and $\arg\{q/p\}$, 
scaled by the appropriate projected event yields, are reported in
Table~\ref{tbl:experimentalreach:golden} for \superb, LHCb and \belletwo.
These are normalized to \babar and \belletwo measurements.
For LHCb, we project 5 \invfb yields from a 37~\invpb exposure
that has $1.2\times 10^5$ $\Dz\to\Kp\Km$ and $15.4\times 10^3$
$\Dz\to\KS\pip\pim$ events
\cite{Spradlin:moriond2011,Wilkinson:dz2hh},
each flavour-tagged by the sign of the pion from $D^*$ decay.  We 
assume that, for the last 4 \invfb, the LHC energy doubles and that 
this doubles the production cross sections.  For the $\KS\pip\pim$ mode we
further assume that a factor 2 in improved trigger efficiency is achieved.

\end{description}

{\bf \underline{Additional precision gained from the \lhcb upgrade}} The
following assumes an upgrade with an integrated luminosity of
50\invfb, starting to take data in 2018. It is assumed that the full
integrated luminosity could be achieved as early as 2030, although
this depends on the performance and upgrade schedule of the \lhc over
the coming decade. Final measurements from the \lhcb upgrade
are expected to appear some years after the \superb ones.
\begin{description}
   \item {\bf $B_s^0$ physics}: The \lhcb upgrade programme may be
     influenced by discoveries or constraints imposed by measurements
     made at \superb and \belletwo as well as the measurement of
     $A_{SL}^s$. In addition, basic exclusive $B_s^0$ branching ratios
     will have been measured by the Super Flavour Factories running at
     $\Upsilon(5S)$ with a high precision, which will help this CERN
     experiment interpret its ratios of branching ratios as absolute
     quantities. As far as $B_s$ golden modes are concerned, an
     upgrade of \lhcb would be able to significantly improve the precision of
     BR($B_s^0 \to \mu^+\mu^-$) from about 30\% to about 8\%, of
     $2\beta_{s}$ from 0.019 to 0.006, and of
     $S(B_s\to\phi\gamma)$ from 0.07 to 0.02~\cite{lhcbupgradeloi}.

    \item {\bf The measurement of $\gamma$}: The ultimate precision of this Unitarity triangle 
    angle for both the \lhcb upgrade and \superb are comparable at the level of $\sim 1^\circ$ for 
    any given channel.  The measurement of the strong phase as a function of the position in the Dalitz 
    plot in $D\to \KS h^+h^-$ decays that is an integral part of the GGSZ method to measure
    $\gamma$ is an input from \superb to the \lhcb upgrade programme. 

   \item {\bf $\sin(2\beta)$ from $J/\psi\KS$}: The \lhcb upgrade is expected to 
     reach a precision of $0.2^\circ$ on $\beta$.  This will provide a useful
     crosscheck of the result from \superb, the latter is expected to have 
     a precision of $0.1^\circ$.

   \item {\bf $\B \to K^{(*)} \mu\mu$}: An upgraded \lhcb experiment
     will select a very large sample, perhaps $~ 100,000$
     events~\cite{WilkinsonElba}, of $\B^0 \to K^{*0} \mu^+\mu^-$
     decays and will be able to do a full angular analysis for this
     channel with high precision. Hence, the precision on quantities measured using this
     particular channel will surpass that of the previously measured
     ones at \superb.  The $\B^0 \to K^{*0} e^+e^-$ channel will have
     perhaps one-tenth of the statistics and will be somewhat smaller
     than the \superb data sample for this channel.  However,
     reconstruction of $B\to X_s\ell^+\ell^-$ in a fully inclusive way
     will not be part of the \lhcb upgrade program.  See
     Section~\ref{sec:experimentalreach} for further comments on this
     channel.

%\superb will not be able to compete with statistics obtained
%     by the \lhcb upgrade on exclusive measurements of this mode.  However, given the results shown
%     in Ref.~\cite{LHCbkstee} it is expected that \superb will be accumulate a larger sample of $\B
%     \to K^{(*)} e^+e^-$ decays, and will therefore make the most precise measurements of
%     $R_{K^{(*)}}$.  Moreover, \superb will be able to exploit the inclusive mode of this decay,
%     $B\to X_s \ell^+\ell^-$, to make stringent tests of the Standard Model.  The inclusive mode is
%     particularly powerful in searching for NP due to the high precision SM calculations available
%     for inclusive observables.  The \lhcb upgrade results will be available many
%     years after the \superb ones.

   \item {\bf Lepton Flavour Violation}: Preliminary studies show that LFV searches by 
     the \lhcb upgrade will not be competitive with \superb~\cite{GuzabProcs,Abe:2007ev,Lees:2010ez}. 
     Anticipated limits from the \lhcb upgrade will be up to an order of magnitude 
     less stringent than \superb or \belletwo.

   \item {\bf Charm mixing}: For the LHCb upgrade, we simply take the
     anticipated factor 10 increase in integrated luminosity at face
     value and assume it will result in a reduction factor $\sqrt{10}$
     in uncertainties reported at the 5 \invfb level.  We realize that
     things may not be that simple in practice since event pile up and
     tracking problems at the higher luminosities may exist.

\end{description}

More details on the \lhcb experiment can be found in Refs.~\cite{Adeva:2009ny,lhcbupgradeloi},
and more information on the proposed \lhcb upgrade can be found in Ref.~\cite{lhcbupgradeloi}.

\subsection{NA62}

The primary goal
of NA62 is a measurement of the branching ratio of $K^+ \to \pi^+ \nu \overline{\nu}$
from a data sample containing 100 signal events. The result of this challenging measurement 
is sensitive to NP in the transition between the second and the first generation.
In the SM it can be used to place a constraint on the CKM parameters,
via a theoretically clean bound on the value of $(\rho, \eta)$, the apex position of 
the unitarity triangle. This additional input can be 
combined with the many different direct and indirect constraints that will be made at \superb
in order to search for NP corrections which could appear in the determination of the unitarity triangle.
This experiment anticipates taking data for two years ($\sim 100$ days of running assuming
a SM branching ratio) in order to accumulate 100 $\pi^+ \nu \overline{\nu}$ events, 
and this measurement should be completed before \superb commences.  The result will provide
information from the kaon sector to constrain models of minimal flavour violation, that will
complement some of the measurements that \superb will make. In addition to the 
$K^+\to \pi^+ \nu \overline{\nu}$ mode, there are a number of other interesting measurements that 
NA62 will make, including a precision measurement of $R_K$, the ratio of $K_{e2}/K_{\mu 2}$ decays,
where $K_{\ell 2} = K^\pm \to \ell^\pm \nu$.  Most of the detectors
should be commissioned for this experiment during a technical run in September 2012. 
Following this, data taking will commence as soon as the LHC injector chain is available
following the 2013/2014 stop\footnote{It is anticipated that the time-scale for this will 
be less than the full shutdown period for the LHC.}.
More details on the NA62 experiment can be found in Ref.~\cite{NA62}, and a recent overview
of the physics can be found in Ref.~\cite{NA62EPS2011}.

\subsection{\koto}

The primary goal of the \koto E14 experiment is the measurement of
BR($K^0_L \to \pi^0 \nu \overline{\nu}$).  This decay is theoretically cleaner
than the corresponding charged mode.  The branching ratio of 
$K^0_L \to \pi^0 \nu \overline{\nu}$ is proportional
to the Wolfenstein parameter $\eta$, and measures the height of the unitarity triangle.
The initial aim of KOTO is to observe this decay, which assuming a SM rate will correspond
to 3 signal events, using data collected during 2012~\cite{kotoichep09}.

\subsection{KLOE 2}

The KLOE 2 experiment at LNF Frascati is another dedicated kaon experiment that will make a number
of important flavour physics measurements.  Some of the goals of this
experiment include testing lepton universality by measuring $R_K$,  and making precision 
tests of \CPT in the neutral kaon system.  In addition, this experiment will be able to make
precision tests of the SM.  There are a number of light meson searches for Dark Forces that 
can be done at KLOE 2, these complement the corresponding exotic spectroscopy studies planned 
at \superb. More information about the diverse KLOE 2 physics programme can be found in 
Refs.~\cite{AmelinoCamelia:2010me}.  KLOE 2 plans to run until the autumn of 2015, and we 
expect this programme to be completed well before \superb acquires its full data sample.

\subsection{\meg, \comet and \mutoe}

The main goal of the \meg, \comet and \mutoe experiments is the search for
LFV in the $\mu \to e$ transition.  In the case of
\meg, the search is for charged LFV via $\mu \to e \gamma$. The
experiment is taking data and reported recently $\text{BR}(\mu\to
e\gamma) < 2.4 \e{-12}$ at 90\%~CL~\cite{meg2011}. In a few years, \meg is
expected to reach its design sensitivity of $10^{-13}$. \comet and
\mutoe are proposed experiments aiming to search for $\mu\to e$
conversion in nuclear material in a time frame that is comparable to
the \superb one. SINDRUM II currently holds the record for the most
precise limit, $R_{\mu e} < 7\cdot 10^{-12}$ (90\% CL), using gold
nuclei~\cite{mu2e:sindrum}. \mutoe is designed to achieve a 90\% CL
upper limit sensitivity of $R_{\mu e} <6\e{-17}$~\cite{Carey:2008zz},
similarly to \comet~\cite{CometProposal:2011}.

The relationship between muon LFV and the $\tau$ LFV searches at
\superb is model dependent. Most NP models have a large
number of free parameters, whose determination requires measurements
of both types of LFV. The degree of correlation between muon and
$\tau$ LFV that is predicted in most analyzed and published NP
scenarios depends on the practical choice to study only minimal
models where most free parameters are unified, or
subjected to symmetries or hierarchies that are not experimentally
constrained, yet are simple, elegant or mimic
experimental patterns in other physics sectors. Therefore, in the
general case it remains essential to experimentally constrain both
the muon and $\tau$ LFV rates.

SUSY models typically predict $\tau\to\mu \gamma \gg \mu\to e \gamma \gg R_{\mu
e}$, with scale factors in the range from $10^{-1}$ to
$10^{-4}$~\cite{Calibbi:2006nq,Arganda:2007jw,Albright:2008ke}.
In most SUSY models, the muon LFV is suppressed if the mixing between the first and
third generation neutrinos is small or zero, i.e.\ when $\theta_{13}$
is small~\cite{Antusch:2006vw}, while the $\tau$ LFV is unaffected.
Global fits give $\theta_{13} < 5.4^\circ$ at 90\%
CL~\cite{Schwetz:2008er}, however recent data
suggest that $\theta_{13}$ might be around the above limit or
higher~\cite{Abe:2011sj}.
In other models like Little Higgs with $T$-parity, the above rates are
either similar or even have an inverted
hierarchy~\cite{Blanke:2009am}. Most published references predict that
\comet and \mutoe are more sensitive than MEG to NP effects, and MEG
is more sensitive than \superb. But even in these minimal NP
frameworks \superb measurements often remain complementary and
essential to validate the models and measure their parameters.

%
% Experimental Reach
%
\renewcommand{\sectionmark}[1]%
                  {\markright{#1}}
\rhead[\fancyplain{}{\bf Experimental Reach}]%
      {\fancyplain{}{\bf\thepage}}

\graphicspath{{ExptReach/}{ExptReach/}}
\section{Experimental Reach}
\label{sec:experimentalreach}

The experimental reach of \superb is considered in two ways, firstly the
precision on observables as discussed in Section~\ref{sec:experimentalreach:comparison} for 
a number of benchmark processes.  Secondly it is important to understand
how to combine the measurements in order to constrain models of 
new physics and the SM.  This is discussed in Section~\ref{sec:experimentalreach:interplay}.
Unless otherwise stated the anticipated precisions for \superb come from
Ref.~\cite{O'Leary:2010af}, those from \belletwo come from Ref.~\cite{Aushev:2010bq}, and those
for LHCb and the LHCb upgrade come from Ref.~\cite{lhcbupgradeloi}.

\subsection{Sensitivity Comparison}
\label{sec:experimentalreach:comparison}

The sensitivity to observables computed from the \superb Golden Modes
is shown in Table~\ref{tbl:experimentalreach:golden}.  In all cases
the quoted precisions are based on the use of the anticipated full
experimental data sets: 75\invab for \superb (+500\invfb at charm
threshold and 1\invab at the $\Upsilon$(5S)), 50\invab for \belletwo,
10\invfb at charm threshold for \besthree, 5\invfb for \lhcb, and
50\invfb for the proposed \lhcb upgrade.  The experiments are listed
in chronological order from left to right, depending on when the full
integrated luminosity will be reached. 

It is evident from the Table that many Golden Modes will only be
measured at \superb and \belletwo.  For those channels, \superb
measurements will generally have smaller statistical uncertainties
than \belletwo measurements.  For measurements whose total uncertainty
dominated by statistics, then, \superb measurements will provide
somewhat greater sensitivity than \belletwo.  However, many results
will be limited by systematic effects, in which cases we would expect
comparable overall precision for \superb and \belletwo.  This issue
highlights an important point for observables that will be
systematically limited at \superb and \belletwo: quoted sensitivities
will depend strongly on any assumptions made when estimating the
systematic uncertainty. In such cases the quoted sensitivities should
be interpreted as a possible range for the ultimate sensitivity
attainable.
A selection of golden modes for other
flavour experiments is given in
Table~\ref{tbl:experimentalreach:goldenotherexpt} (this table excludes
entries corresponding to \superb golden modes).

\begin{widetext}
\begin{table*}[!h]
\caption{Experimental sensitivities for \superb Golden Modes.  Where appropriate, the sensitivity for other
experiments is also indicated.  
The current state of the art is also shown (usually an average of \babar and \belle results).
Entries marked with (est.) for \belletwo are estimated from
the \superb results, scaling by the difference in integrated luminosity.}
\label{tbl:experimentalreach:golden}
\begin{center}
\small
\begin{tabular}{l||c||c|c|c|c||c}\hline
Observable/mode & Current & \lhcb & \superb & \belletwo &  \lhcb upgrade & theory\\ 
  & now & (2017) & (2021) &(2021) &  (10 years of  & now \\ 
  &     & $5\invfb$ & $75\invab$ & $50\invab$ & running) $50\invfb$ & \\ \hline
\multicolumn{7}{c}{$\tau$ Decays}\\ \hline
$\tau\to \mu\gamma$ ($\times 10^{-9}$)     & $<44$ & & $<2.4$ & $<5.0$ &  \\
$\tau\to e\gamma$ ($\times 10^{-9}$)       & $<33$ & & $<3.0$ & $<3.7$ (est.) &  \\
$\tau\to \ell\ell\ell$ ($\times 10^{-10}$) & $<150 - 270$ & $<244$
   \footnote{Based on extrapolation of the results presented in~\cite{lhcbtauphysics} for $2\invfb$ to $5\invfb$.}
   & $<2.3-8.2$ & $<10$  & $<24$ \footnote{Extrapolation from \lhcb assuming no background.} & \\
%CP Violation in $\tau^+ \to \KS\pi^+\nu$   & X & X & Y & n.a. & n.a. & n.a. \\

\hline
\multicolumn{7}{c}{$B_{u,d}$ Decays}\\ 
\hline

BR($B\to\tau \nu$) ($\times 10^{-4}$) & $1.64 \pm 0.34$ & & 0.05 & 0.04 & & $ 1.1 \pm 0.2 $\\
BR($B\to\mu \nu$) ($\times 10^{-6}$)  & $<1.0$ & & 0.02 & 0.03 & & $0.47 \pm 0.08$ \\
BR($B\to K^{*+}\nu \overline{\nu}$) ($\times 10^{-6}$) & $<80$ & & $1.1$ & $2.0$ &  & $6.8 \pm 1.1$\\
BR($B\to K^{+}\nu \overline{\nu}$) ($\times 10^{-6}$)  & $<160$ & & $0.7$ & $1.6$ &  & $3.6 \pm 0.5$\\
BR($B\to X_s \gamma$) ($\times 10^{-4}$) & $3.55 \pm 0.26$ &  & 0.11 & 0.13 &  0.23 & $3.15\pm0.23$ \\
$A_{CP}(B\to X_{(s+d)} \gamma)$ & $0.060\pm 0.060$ & & 0.02 & 0.02 & & $\sim 10^{-6}$ \\  
%%% K*mumu
$B\to K^{*} \mu^+\mu^-$ (events)   & 250\footnote{The separate yields for $B\to
  K^{*}\ell^+\ell^-$ in muon and electron channels are not published
  for all experiments. We assume an equal number of selected
  events in the two channels for SuperB and Belle II.} & 8000 & 10-15k\footnote{The ranges presented represent results of independent extrapolations
of \babar and \belle data.} & 7-10k & 100,000 & - \\
BR($B\to K^{*} \mu^+\mu^-$)  ($\times 10^{-6})$   & $1.15 \pm 0.16$ &  & 0.06  & 0.07 & & $1.19 \pm 0.39$ \\
%%% K*e+e-
$B\to K^{*} e^+e^-$ (events)       & 165 & 400
     %\footnote{Based on extrapolation of the results presented 
     %in~\cite{LHCbkstee} for $2\invfb$ to $5\invfb$. Earlier studies exist, but do not include 
     %trigger efficiencies which are significant for this mode. } 
     & 10-15k & 7-10k &
5,000
% jw - I think this footnote was placed here by mistake. 
%\footnote{Extrapolation from \lhcb assuming no background.} 
& - \\
BR($B\to K^{*} e^+e^-$)  ($\times 10^{-6})$   & $1.09 \pm 0.17$ &  & 0.05 & 0.07 & & $1.19 \pm 0.39$ \\
$ A_{FB}(B\to K^{*} \ell^+\ell^-)$ & $0.27 \pm 0.14$\footnote{For $ 0 < q^2 < 4.3 $ GeV$^2$, although we include the \babar measurement made in the $q^2$ range 0-6.25 GeV$^2$.} & \footnote{Please see discussion of $b\to s\ell^+\ell^-$ in text.} & 0.040 & 0.03 & & $-0.089 \pm 0.020$ \\   
$B\to X_s \ell^+\ell^-$ (events) & 280 & & 8,600 & 7,000 & & - \\
BR($B\to X_s \ell^+\ell^-$) ($\times 10^{-6}$)\footnote{For low-$q^2$ region, 1-6 GeV$^2$.}  & $3.66 \pm 0.77$\footnote{Current measurement is over
full $q^2$-range.} &  & 0.08 & 0.10 &  & $1.59\pm0.11$ \\
$S$ in $B\to \KS\pi^0\gamma$   & $-0.15 \pm 0.20$ &  & 0.03 & 0.03 & & -0.1 to 0.1\\
$S$ in $B\to \eta^\prime K^0$  & $0.59 \pm 0.07$ & & 0.01 & 0.02 &  & $\pm 0.015$\\
$S$ in $B\to \phi K^0$         & $0.56 \pm 0.17$ & 0.15 & 0.02 & 0.03 & 0.03 & $\pm 0.02$\\

\hline
\multicolumn{7}{c}{$B_s^0$ Decays}\\ 
\hline
BR($B_s^0 \to \gamma\gamma$) ($\times 10^{-6}$) & $<8.7$ & & 0.3 & $0.2 -0.3$& & 0.4 - 1.0 \\
$A_{SL}^s$ ($\times 10^{-3}$) & $-7.87 \pm 1.96$ \footnote{The semileptonic asymmetry $A_{SL}^b$ for a mixture of all beauty hadron decays is shown. The D0 measurement 
%\cite{Abazov:2010hj}
\cite{Abazov:2011yk} is not yet confirmed by other experiments.}  & 
  \footnote{Monte Carlo studies have been performed for \lhcb \cite{lhcbasym}.  While statistically 
   competitive it should be noted that systematic uncertainties will be important, and can be 
   controlled better by measuring differences of the asymmetry in $B_s$ and $B_d$ decays.} & 4. & 5. (est.) & & $0.02 \pm 0.01$ \\
\hline
\multicolumn{7}{c}{$D$ Decays}\\ 
\hline
$x$        & $(0.63 \pm 0.20 $\% & 0.06\%  & 0.02\%  & 0.04\%  & 0.02\%  &  
$\sim~10^{-2}$
\footnote{SM theoretical estimates for charm mixing parameters are
difficult due to the predominance of long range effects.  It is generally
accepted that $x$ and $y$ in the 1\% range, compatible with current
experimental measurements, can be accommodated and that \CPV asymmetries
are likely to be $\sim 10^{-3}$.  
See \cite{Falk:2001hx,FALK2,Kreps:2011cb,Bobrowski:2010xg,Petrov:2011un} and earlier
references for more details.}  \\
$y$        & $(0.75 \pm 0.12)$\% & 0.03\%  & 0.01\%  & 0.03\%  & 0.01\%  & 
$\sim~10^{-2}$ (see above). \\
$y_{CP}$   & $(1.11 \pm 0.22)$\% & 0.02\%  & 0.03\%  & 0.05\%  & 0.01\%  &
$\sim~10^{-2}$ (see above). \\
$|q/p|$    & $(0.91 \pm 0.17)$\% & 8.5\%   & 2.7\%   & 3.0\%   & 3\%     &
$\sim~10^{-3}$ (see above). \\
$\arg\{q/p\}$ ($^{\circ}$)
           & $-10.2 \pm 9.2$ & 4.4\dgr & 1.4\dgr & 1.4\dgr & 2.0\dgr &
$\sim~10^{-3}$ (see above). \\
\hline
\multicolumn{7}{c}{Other processes Decays}\\ 
\hline
$\sin^2\theta_W$ at $\sqrt{s} = 10.58 \gevcc$ & && 0.0002 & \footnote{The \belletwo experiment can be used to measure axial vector couplings related to $\sin^2\theta_W$, but is unable to perform a stand-alone measurement of the weak mixing angle.} & & clean \\
\hline
\end{tabular}
\end{center}
\end{table*}
\end{widetext}

\begin{table*}[!h]
\caption{Some of the golden modes to be measured at facilities other than \superb.}
\label{tbl:experimentalreach:goldenotherexpt}
\begin{center}
\small
\begin{tabular}{l||c||c|c}\hline
Observable & Current value & Experiment & Precision \\ \hline
BR($B_s \to\mu\mu$) ($\times 10^{-9}$)    & $<11$\footnote{Combined LHC limit, CMS-PAS-BPH-11-019; LHCb-CONF-2011-047; CERN-LHCb-CONF-2011-047.} 
                                                  & \lhcb & $\pm 1$ \\
                                          &       & \lhcb upgrade & $\pm 0.3$ \\
$2\beta_{s}$
from $B_s^0 \to J/\psi \phi$ (rad)& $0.13 \pm 0.19$\footnote{There is a second solution at $2\beta_{s} \sim -3.2$, See G. Raven's contribution to Lepton Photon 2011.  It will be possible to resolve this ambiguity in the near future.}
                                                  & \lhcb & 0.019 \\
                                          &       & \lhcb upgrade & 0.006 \\
$S$ in $B_s\to \phi \gamma$               &       & \lhcb  & 0.07 \\
                                          &       & \lhcb upgrade & 0.02 \\
$K^+\to \pi^+ \nu \overline{\nu}$ (\% BR measurement)   & 7 events & NA62 & 100 events (10\%)\\
$K^0_L \to \piz \nu \overline{\nu}$       &       & KOTO & 3 events (observe) \\
$BR(\mu\to e\gamma)$ ($\times 10^{-13}$)  & $<280 $  & MEG        & $<1$ \\
$R_{\mu e}$                               & $<7\times 10^{-12}$     & COMET/Mu2E & $<6\times 10^{-17}$ \\ 
\hline
\end{tabular}
\end{center}
\end{table*}

While Table~\ref{tbl:experimentalreach:golden} gives an idea of the
relative capabilities of the different flavour experiments, it is
necessarily simple for certain measurements. In particular, the
measurements made in the $b\to s\ell^+\ell^-$ sector are too numerous
and complex to describe well in table format.  Both exclusive modes
(e.g. $B\to K^{(*)}\ell^+\ell^-$) and inclusive modes ($B\to
X_s\ell^+\ell^-$) are of interest in searching for NP, indeed the
theoretical uncertainties associated with the inclusive modes are
likely to be smaller than those of the exclusive
modes~\cite{O'Leary:2010af}, thus increasing the sensitivity to NP
using inclusive measurements.  Furthermore, final states with both
muons and electrons are interesting to measure separately.  For each
specific channel, then, there are several observables that one can
measure: branching ratios, direct \CP asymmetries, isospin
asymmetries and the ratio of the rate of the muon channel to the rate
of the electron channel.  Additionally, there are observables that
depend on the angular distributions in $B\to K^*\ell^+\ell^-$ and
$\B\to X_s\ell^+\ell^-$: the forward backward lepton asymmetry, the
$K^*$ polarization fraction and other asymmetries, 
such as $A_T^{(2)}$ and related variables~\cite{bsll:angular_variables},
all of which are sensitive to NP.

\lhcb (and its possible upgrade) will make its greatest contribution
in the exclusive mode $B^0\to K^{*0} \mu^+\mu^-$, for which it will
accumulate high statistics, as reported in
Table~\ref{tbl:experimentalreach:golden}.  \superb will perform
complementary measurements by accumulating large samples of the
related modes $B^+\to K^{*+}\mu^+\mu^-$ and $B^0\to K^{*0} e^+e^-$,
which will permit the study of several quantities that are sensitive
to NP, such as the $e$-to-$\mu$ ratio,
$R_{K^*}$.  \superb, in contrast to \lhcb, will fully explore the
$b\to s\ell^+\ell^-$ sector, with measurements of all relevant
observables with high statistics in all channels, including the
inclusive modes. Estimates of the reach of \superb for many of these
observables can be found in~\cite{O'Leary:2010af}. \belletwo will make
similar measurements, with generally somewhat lower statistics.

The precisions of observables computed from the CKM measurement modes
are shown in Table~\ref{tbl:experimentalreach:calibration}.  Regarding
the three angles of the Unitarity Triangle, one can see that the most
precise measurements of $\alpha$ will come from \superb and \belletwo,
and that for the angles $\beta$ and $\gamma$ it is possible to obtain
results of comparable precision from \superb, \belletwo and the \lhcb
upgrade.  The current experimental situation with regard to \Vub and
\Vcb is that inclusive and exclusive measurements are not in good
agreement.  The additional data available at \superb will allow a
precision comparison of these quantities.  It should be noted that for
the time-dependent \CP analyses performed by \babar, one typically (i)
includes more final states, and (ii) performs a more sophisticated
analysis in order to obtain a smaller uncertainty than the
corresponding \belle results.  The main difference between the
extrapolations for \superb and \belletwo is that the latter has larger
{\em irreducible} systematic uncertainties based on experience with
the silicon vertex detector capabilities from \belle.  It is
reasonable to presume that \belletwo will perform better than expected
with the proposed pixel detector for the innermost layer and a five
layer double-sided SVT akin to the \superb design.  For these reasons
we believe the \belletwo predictions to be slightly underestimating
their capability.  It is clear from this table, that state of the art
for this set of observables, as defined by \superb and \belletwo circa
the end of 2021, will not be surpassed by any additional input from an
\lhcb upgrade.  Including results from the \lhcb upgrade will have
only marginal impact and may improve the precision of the world
averages of $\beta$ and $\gamma$ measurements by 30\% when those
results become available once the full \lhcb upgrade data sample has
been analyzed.  Only \superb and \belletwo will be able to measure the
angle $\alpha$ with precision.

In summary, it is clear that \superb has the best expected sensitivities
(in some cases, together with \belletwo) for virtually all the
golden and precision CKM channels that it aims to measure.  In terms of
performing direct and indirect constraints on the SM, only \superb
will be able to measure a complete set of constraints in each of these
categories in order to push our current 10\% determination of the CKM
mechanism, down to the 1\% level using angles.  The stated improvements
on \Vub and \Vcb include a reasonable level of progress for the precision
of Lattice QCD inputs. While \belletwo will match
sensitivities of \superb in some measurements, \superb will be able to make
additional measurements owing to its charm threshold program and beam
polarisation. In time the \lhcb upgrade should be able to improve
constraints on $B^0\to K^{*0}\mu^+\mu^-$ and in the charm mixing
parameter $y_{CP}$.

\begin{table*}[!h]
\caption{Experimental sensitivities for \superb precision CKM
  measurement modes.  Where appropriate, the sensitivity for other
  experiments is also indicated.  The current state of the art is also
  shown, with $\alpha$ and $\gamma$ taken from
  UTFit~\cite{Bona:2005eu}, and the remainder of the observables taken
  from HFAG.  Entries marked with (est.) for \belletwo are estimated
  from the \superb results, scaling by the difference in integrated
  luminosity.  In order to control theoretical uncertainties on the
  charmonium $\beta$ measurement, one has to be able to measure
  $B_d\to J/\psi \piz$ or $B_s\to J/\psi \KS$.  The precision for the
  former has been determined for \superb integrated luminosities,
  however while \lhcb will be able to measure the latter, the
  precision on this control channel is not yet know, and is indicated
  by a `?'.}
\label{tbl:experimentalreach:calibration}
\begin{center}
\small
\begin{tabular}{l|c|c|c|c|c|c}\hline
Observable/mode & Current & \lhcb & \superb & \belletwo  &  \lhcb upgrade   & theory\\  
                & now     & (2017)   & (2021) &(2021) & (10 years of running) & now \\ \hline 
                &         & $5\invfb$ & $75\invab$ & $50\invab$ & $50\invfb$ \\ \hline
$\alpha$ from $u\overline{u}d$ & $6.1^\circ$  & $5^\circ$\footnote{This estimate is based
on the study in Ref.~\cite{thesis:Robert}, combined with the expectation that the $\B\to \pi\pi\pi$ approach will be
systematics limited at this level~\cite{Deschampspc}.  Ignoring systematic errors one may reach a precision of $3^\circ$ with
10\invfb of data.}
 & $1^\circ$   & $1^\circ$    & \footnote{It is not clear how well any \lhcb upgrade might be able to measure this angle, however it is unlikely to be competitive with \superb.} & $~1-2^\circ$\\
$\beta$ from $c\overline{c}s$ (S) & $0.8^\circ$ (0.020) & $0.5^\circ$ (0.008) & $0.1^\circ$ (0.002) & $0.3^\circ$ (0.007) &  $0.2^\circ$ (0.003) & clean\\
$\,\,\,$ $S$ from $B_d\to J/\psi \piz$  & 0.21 & & 0.014 & 0.021 (est.) & & clean\\ 
$\,\,\,$ $S$ from $B_s\to J/\psi \KS$   &      & ? & & & ? & clean \\
$\gamma$ from $B\to DK$        & $11^\circ$   &  $\sim 4^\circ$ & $1^\circ$   & $1.5^\circ$   & $0.9^\circ$ & clean\\ \hline
\Vcb (inclusive) \%& 1.7 & &0.5\% & 0.6 (est.)   &  & dominant \\
\Vcb (exclusive) \%& 2.2 & &1.0\% & 1.2 (est.)   &  & dominant \\
\Vub (inclusive) \%& 4.4 & &2.0\% & 3.0 &    & dominant \\
\Vub (exclusive) \%& 7.0 & &3.0\% & 5.0 &    & dominant \\ 
\hline
\end{tabular}
\end{center}
\end{table*}

\subsection{Interplay between observables}
\label{sec:experimentalreach:interplay}

The interplay between measurements of golden modes discussed in this document and
a sample of NP models is highlighted by the golden matrix of
Table~\ref{tbl:experimentalreach:goldenmatrix}.
The pattern of deviations from the SM (\threestar = large, \twostar = medium, 
\onestar = observable, but small) and their correlations can be used to identify the
structure of the NP Lagrangian.

\begin{figure}[!ht]
  \begin{center}
  \includegraphics[width=0.45\textwidth]{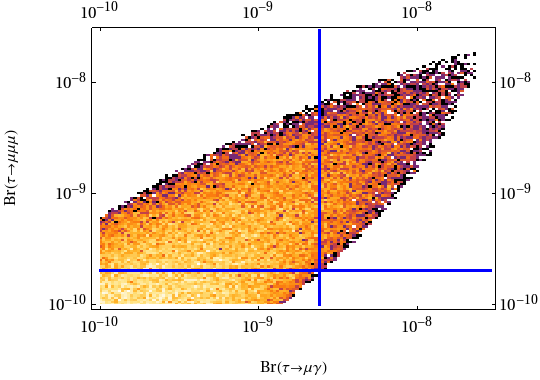}
  \caption{\label{fig:LHT-FV}
           Correlation between BR($\tau\to\mu\gamma$) and BR($\tau\to\mu\mu\mu$) in the LHT
           for the symmetry breaking scale $f=500$ GeV. 
           The solid lines indicate expected \superb sensitivities.
           }
  \end{center}
\end{figure}

An example of this can be seen with regard to the interpretation of 
LFV $\tau$ decays: in models where the LFV amplitude is dominated by the electric
dipole operator, the ratio of the rates of $\tau\to 3\ell$ and
$\tau\to\mu \gamma$ is governed by the electromagnetic coupling $\alpha_e$.
In other models, this ratio could become of order one, a very favourable situation for
\superb which could observe the decay $\tau\to 3\ell$ only.
In such a case, many models, including the MSSM, would be ruled out.
The Littlest Higgs model with T parity (LHT) is an example of a model which could
produce large ratios of the rates of $\tau\to 3\ell$ and $\tau\to\mu \gamma$.
Figure~\ref{fig:LHT-FV} (from Ref.~\cite{Blanke:2009am}) illustrates the 
correlation between these two golden observables
in the context of the LHT model with a symmetry breaking scale $f=500$ GeV.

\begin{sidewaystable*}
\caption{Golden matrix of observables/modes that can be measured at \superb.
The size of the NP effect in a given model/scenario is indicated by the number of stars:
\threestar, \twostar, \onestar. The more stars the larger the effect.
Additional notes: CKM indicates precision CKM required.
Lattice QCD improvements are required at the predicted level (See CDR/White paper for details).
A question mark indicates that the channel has not been studied yet.
The information here is taken from previous work done within \superb and from
Refs.~\cite{Blanke:2009am,Altmannshofer:2009ne,Girrbach:2011an} where details on
specific models can be found.}
\label{tbl:experimentalreach:goldenmatrix}
\begin{center}
\small
\begin{tabular}{|l|c|c|c|c|c||c|c|c|c|c|c|c|}\hline
Observable/mode & charged Higgs    & MFV NP         & non-MFV NP & NP in        & Right-handed  & LHT & \multicolumn{6}{|c|}{SUSY}\\
                & high $\tan\beta$ & low $\tan\beta$    &  2-3 sector      & $Z$ penguins & currents      &     & AC & RVV2 & AKM & $\delta LL$ & FBMSSM & GUT-CMM\\
\hline
$\tau\to \mu\gamma$                  & & & & & & & \threestar &  \threestar &  \onestar &  \threestar & \threestar & \threestar \\
$\tau\to \ell\ell\ell$               & & & & & & \threestar & & & & & & ? \\
\hline
$B\to \tau \nu, \mu\nu$              & \threestar (CKM) & & & & & & & & & & & \\
$B\to K^{(*)+}\nu \overline{\nu}$    & & & \onestar & \threestar & & & \onestar & \onestar & \onestar& \onestar & \onestar & ? \\
%$\alpha$                             & & & & & & & & & & & \\
%$\beta$                              & & & & & & & & & & & \\
%$\gamma$                             & & & & & & & & & & & \\
$S$ in $B\to \KS\pi^0\gamma$         & & & \twostar & & \threestar & & & & & & & \\
$S$ in other penguin modes           & & & \threestar (CKM) & & \threestar & & \threestar & \twostar & \onestar & \threestar & \threestar & ? \\
$A_{CP}(B\to X_s \gamma)$            & & & \threestar & & \twostar & & \onestar & \onestar & \onestar & \threestar & \threestar & ? \\
$BR(B\to X_s \gamma)$                & & \onestar & \twostar & & \onestar & & & & & & & \twostar\\
$BR(B\to X_s \ell \ell)$             & & & \twostar & \onestar & \onestar & & & & & & & ? \\
$B\to K^{(*)} \ell \ell$ (FB Asym)   & & & & & & & \onestar & \onestar & \onestar & \threestar & \threestar & ? \\
\hline
%$B_s \to\mu\mu$                      & & & & & & & \threestar &  \threestar &  \threestar &  \threestar &  \threestar & \onestar\\
%$\beta_s$ from $B_s \to J/\psi \phi$ & & & & & & & \threestar & \threestar &\threestar  & \onestar & \onestar & \threestar\\
%$B_s \to \gamma\gamma$               & & & & & & & & & & & \\
$a_{sl}^s$                           & & & \threestar & & & \threestar & & & & & &\threestar\\
\hline
Charm mixing                         & & & & & & & \threestar & \onestar & \onestar & \onestar & \onestar & \\
CPV in Charm                         & \twostar & & & & & & & & & \threestar & & \\
\hline
%$\sin^2\theta_W$ at \FourS           & & & & & & & & & & & \\
%$\sin^2\theta_W$ at Z-pole           & & & & & & & & & & & \\ \hline
\end{tabular}
\end{center}
\end{sidewaystable*}

\begin{figure}[!ht]
  \begin{center}
  \includegraphics[width=0.45\textwidth]{figures/epseta-wp.pdf}
  \includegraphics[width=0.45\textwidth]{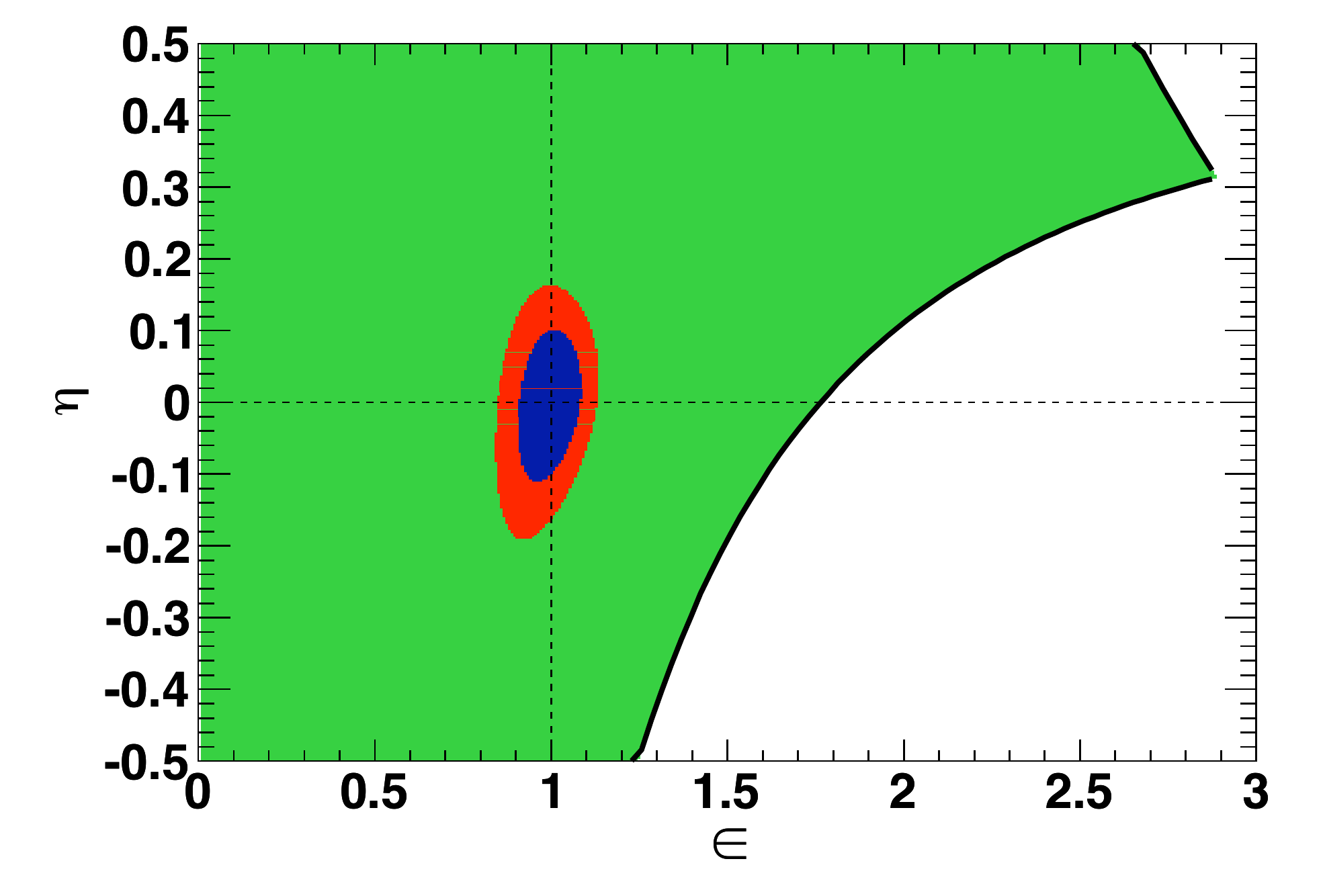}
  \caption{\label{fig:angular_constraint} The constraint on the
    $(\epsilon,\eta)$ plane obtained from $B^0 \to K^{*0} \nu
    \overline \nu$ decays. The parameters are defined in
    Ref.~\cite{Altmannshofer:2009ma} (top figure taken from this
    reference). The top plot shows the dependence of $\epsilon$ and
    $\eta$ on the measurable quantities.  The horizontal band
    corresponds to the constraint from a polarisation measurement in
    $B \to K^{*} \nu\overline \nu$, and the band increasing
    (decreasing) from left to right corresponds to $B\to K
    \nu\overline \nu$ ($B\to K^{*} \nu\overline \nu$).  The vertical
    constraint comes from the inclusive measurement of $B\to
    X_s\nu\overline \nu$.  The error bands shown in this plot are
    theoretical.  The oval contours of the bottom plot show the expected
    experimental constraint using measurements of the branching ratios
    of $B \to K^{(*)} \nu \overline \nu$ and the angular analysis of
    $B^0 \to K^{*0} \nu \overline \nu$ with $75\,ab^{-1}$ at \superb.
    The green region represents the constraint based on current
    experimental limits.}
  \end{center}
\end{figure}

Flavor changing neutral current processes can provide evidence for NP.
For example one can compare the results of experimental measurements
in $B\to K^{(*)} \nu\overline{\nu}$ decays with the SM expectation to
elucidate information on models with right handed
currents. Figure~\ref{fig:angular_constraint} shows the constraint on
the model-independent parameters $(\epsilon,\eta)$, defined in
Ref.~\cite{Altmannshofer:2009ma}.  The SM expectation for these
parameters is $(1, 0)$.  The polarisation measurement in the $K^*$
mode starts to become important once datasets of 75\invab have been
achieved.  Other observables that can be used to test for right handed
currents include the time-dependent \CP asymmetry parameters measured
in $B^0\to \KS\piz\gamma$, a mode which is only accessible in an \epem
environment.

\begin{figure}[!ht]
  \begin{center}
   \includegraphics[width=0.46\textwidth]{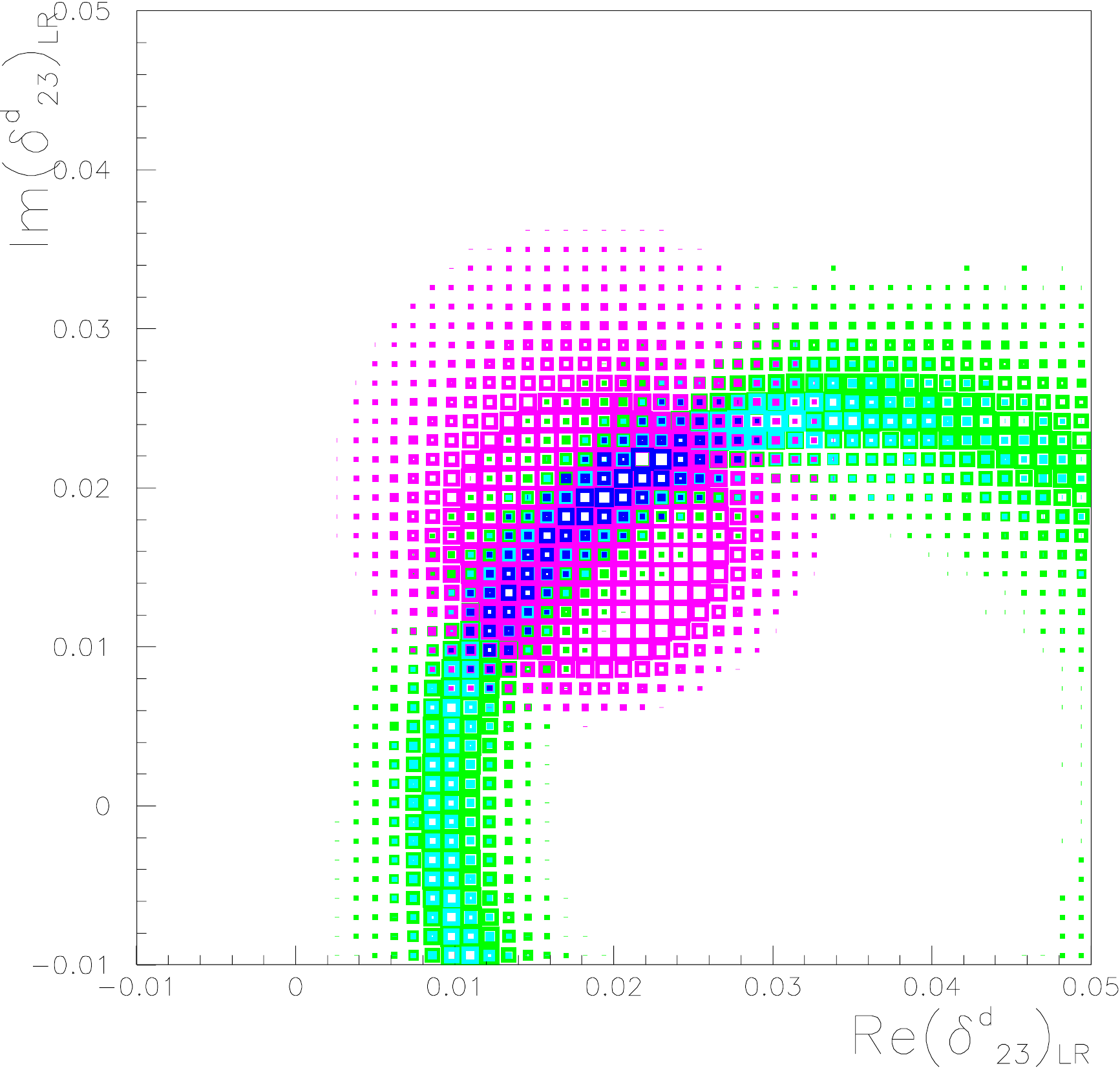}
   \includegraphics[width=0.51\textwidth]{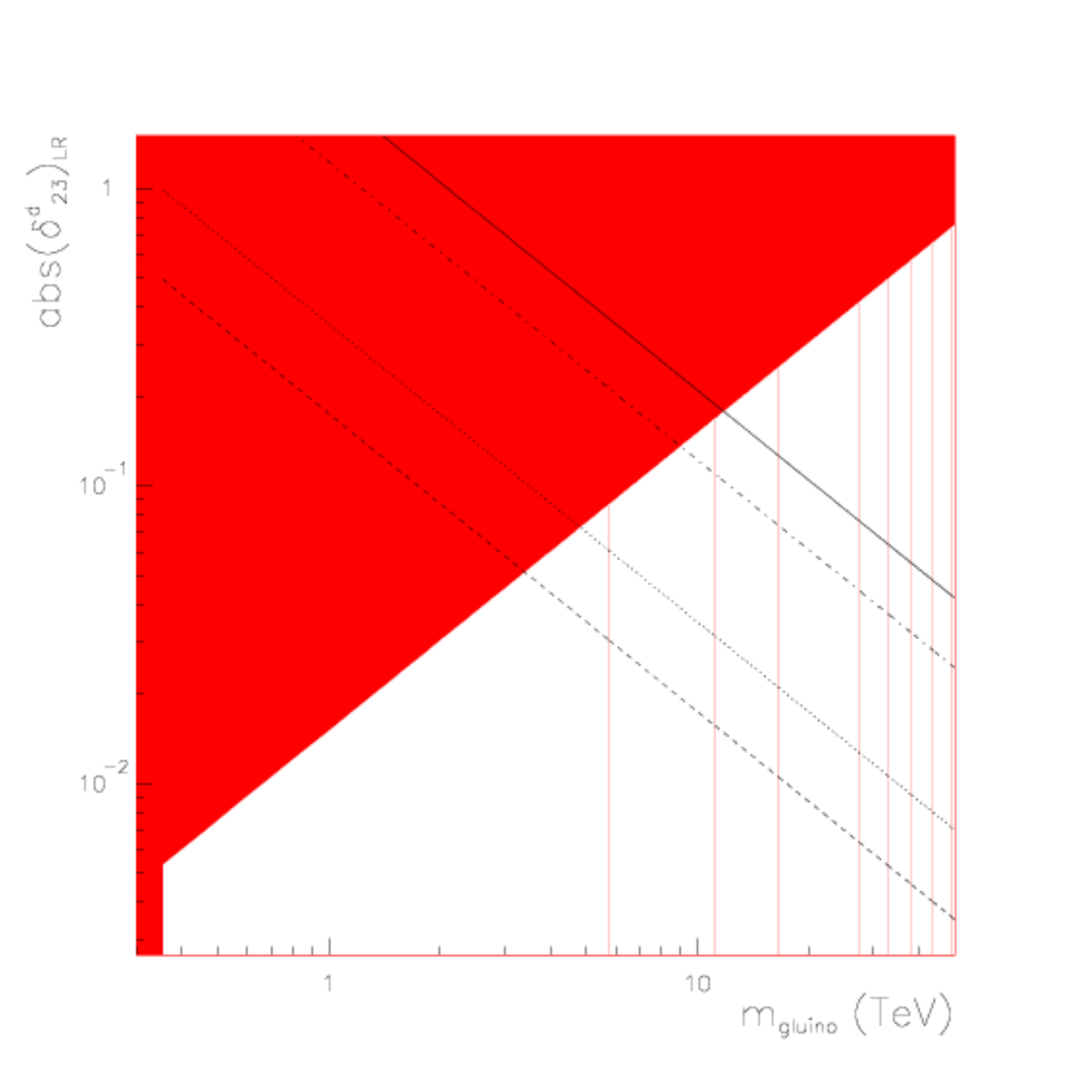}
   \caption{\label{fig:MI}
           Top plot: extraction of $(\delta_{23}^d)_{LR}$ from the measurements
           of $a_\mathrm{CP}(B_d\to X_s\gamma)$ (magenta), $BR(B_d\to X_s\gamma)$ (green) and
           $BR(B_d\to X_s\ell^+\ell^-)$ (cyan) with the errors expected at next-generation flavor
           experiments. Central values are generated using $(\delta_{23}^d)_{LR}=0.028\,e^{i\pi/4}$
           and squark and gluino masses at $1$ TeV. Bottom plot: region of the parameter space
           where a non-vanishing $(\delta_{23}^d)_{LR}$ can be extracted with at least $3\sigma$
           significance (in red).
           }
  \end{center}
\end{figure}

The set of inclusive observables encompassing $b\to s\gamma$ and $b\to s \ell\ell$ transitions
can be used to constrain the MSSM with generic soft SUSY-breaking terms.  These measurements
are essentially constraints on off-diagonal entries of the squark mixing matrices as
functions of the average SUSY mass. Figure~\ref{fig:MI} shows the 
constraints that can be achieved on the real and imaginary parts of the mass insertion 
parameter $(\delta_{23}^d)_{LR}$ for a SUSY scale ~1 TeV (top plot) and the region of mass insertions 
and SUSY masses where a deviation from the SM larger than 3$\sigma$ can be observed at
\superb (bottom plot).  SUSY searches at the \lhc are starting to rule out the parameter space
for low energy scales $\sim 1$ TeV, and it is possible that by the time \superb
starts running, there will be no evidence for new physics.  In this case
combinations of flavour observables, such as those described here, can provide
a window to probe to energies beyond the reach of the \lhc. For mass insertions $\sim 0.1$ the energy 
scale probed by \superb is 10 TeV, and for larger values the scale increases.
%Conversely, if something is found at the \lhc, visible effects are expected at \superb
%for mass insertions as small as $0.01$.
Direct searches for SUSY at the \lhc place the energy scale above $\sim 1$ TeV.  
As a result we expect mass insertions to give visible effects for magnitudes 
greater than 0.01.

\begin{figure}[!ht]
  \begin{center}
  \includegraphics[width=0.45\textwidth]{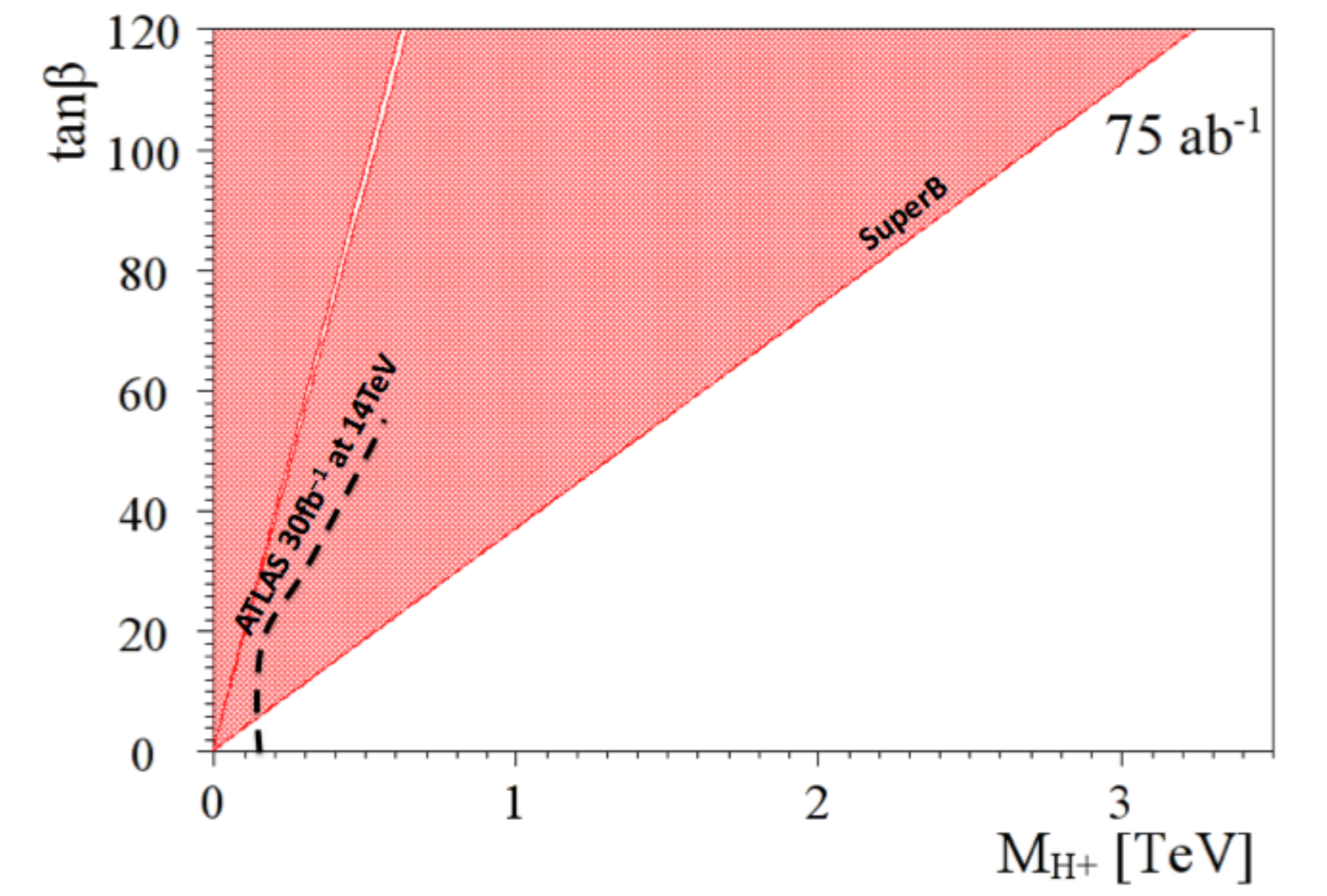}
  \caption{\label{fig:btaunu}
The constraint on the mass of a charged Higgs boson as a function of $\tan\beta$ in a
2HDM-II (95\% C.L.). The constraint anticipated from the LHC with 30\invfb of data
collected at a centre of mass
energy of 14 TeV is also shown for comparison.  The \superb constraint is dominated
by $B\to \tau \nu$ up to luminosities $\sim 30\invab$, and the $B\to \mu \nu$
contribution dominates above this.  The ATLAS constraint is taken from arXiv:0901.0512.}
  \end{center}
\end{figure}

It is also possible to constrain the mass of a charged Higgs boson in Two-Higgs Doublet Models Type II 
(2HDM-II) at \superb. The constraint one obtains from the measurement of $B\to \ell \nu$
decays on the mass of a $H^+$ boson as a function of $\tan\beta$ is given in
Figure~\ref{fig:btaunu}. Similar constraints can be obtained for the MSSM.
One can see that the excluded region obtained
from the \superb measurements is far greater than that obtained from direct searches at the
LHC, assuming a data sample of 30\invfb of data collected at a centre of mass
energy of 14 TeV by ATLAS. In the MSSM, stronger bounds on the same plane could be obtained from
BR$(B_s\to\mu^+\mu^-)$ measured at \lhcb depending on the value of other SUSY parameters.

\begin{figure}[!ht]
  \begin{center}
  \includegraphics[width=0.45\textwidth]{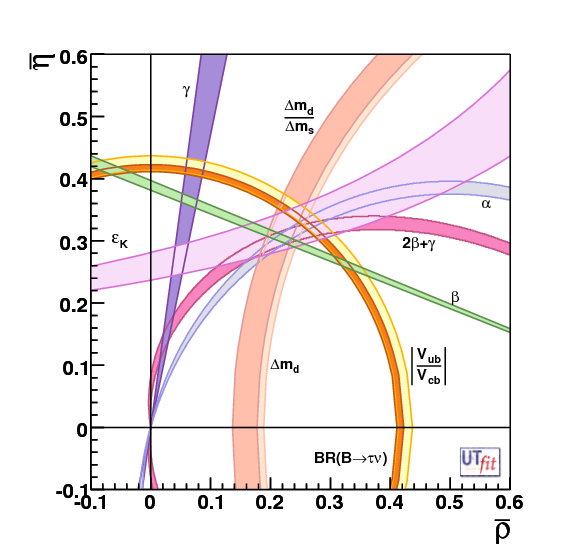}
  \caption{\label{fig:precisionCKM} 
           Constraints on the $(\bar \rho, \bar \eta)$ plane using measurements from \superb.
           This is the dream scenario, where existing central
	   values are extrapolated to sensitivities expected from \superb with $75\,ab^{-1}$
           and one can see that the constraints
           are not consistent with the CKM scenario.
           This highlights the importance of performing a precision CKM test with \superb.}
  \end{center}
\end{figure}

Finally in the context of precision tests of the CKM matrix, \superb would be able to reduce
the constraints on the $\rho$, $\eta$ plane to the level illustrated by Figure~\ref{fig:precisionCKM}.
The central values assumed correspond to the current measurements.  If one extrapolates to the 
precision attained with 75\invab of data from \superb, a percent level test of CKM could be 
performed, and if we assume for the sake of illustration that the central values would stay at
today's values, then  we would obtain a clear NP signal.

%
% Summary
%
\renewcommand{\sectionmark}[1]%
                  {\markright{#1}}
\rhead[\fancyplain{}{\bf Summary}]%
      {\fancyplain{}{\bf\thepage}}

\graphicspath{{Summary/}{Summary/}}
\section{Summary}
\label{sec:summary}

The high energy physics community has a strong programme of existing and planned
flavour physics experiments.  These experiments complement each other well, and
if one combines all of the results, then one will be able to restrict the 
parameter space and viable classes of NP. The flavour landscape when the 
first physics results from \superb are available in 2017 will have been 
defined by the $B$ Factories (\babar and \belle), as well as \besthree, 
\lhcb, and the kaon experiments.  The two \sffs, 
\superb and \belletwo, will dominate the redefinition of this landscape 
between 2017 and 2021 as they will collect two orders of magnitude more data
than their predecessors.  This redefinition will have many significant 
contributions from \superb, which will collect more data than \belletwo 
and has the benefit of two unique features: polarised electrons and 
the ability to collect data near charm threshold.  The \lhcb upgrade
will improve upon the results from \lhcb, integrating 50\invfb by
the end of its lifetime.  Each of these areas has unique potential to 
teach us something new about nature.

In order to maximize our understanding of new physics, one has to measure as many golden
channels as possible, and determine the correlations of any deviations from SM expectations
that would arise from NP. {\em A priori,} one can not predict the model to pursue, 
and thus we arrive at a golden matrix of observables vs. viable NP scenarios. 
The measurement of these observables would contribute to the long-term program of
identifying and reconstructing the NP Lagrangian. To this end, a key ingredient could come from
direct searches at the \lhc, in case new particles are found before the start of \superb.
Yet, even if no new particle is unearthed, measurements from \superb could still
be sensitive to scales larger than that achievable at the \lhc.  Some of the indirect
constraints that will be made at the \sffs extend the search capabilities beyond the 
\lhc, for example in the case of a charged Higgs boson, where current experimental bounds 
from \babar and \belle are equivalent to those expected from ATLAS with 30\invfb of data at
14 TeV collision energies.  The golden matrix is shown in 
Table~\ref{tbl:experimentalreach:goldenmatrix}, with expected numerical precisions indicated
in Tables~\ref{tbl:experimentalreach:golden} and \ref{tbl:experimentalreach:goldenotherexpt}
for the \superb and a selection golden modes from other experiments, respectively.

\bibliography{comparison}{}

\end{document}